\documentclass[aps,prd,twocolumn,superscriptaddress]{revtex4-2}
\usepackage{bm}
\usepackage{amsfonts}
\usepackage{amssymb}
\usepackage{latexsym}
\usepackage[utf8]{inputenc}
\usepackage{graphicx}
\usepackage{amsmath}
\usepackage{mathpazo}
\usepackage{textcomp}
\usepackage{float}
\usepackage{booktabs}
\usepackage{ragged2e}
\usepackage{hyperref}
\hypersetup{colorlinks,citecolor=blue}
\hypersetup{colorlinks=true,linkcolor=red,filecolor=magenta,urlcolor=blue}
\usepackage{xcolor}
\newcommand{\rev}[1]{#1}

\newcommand{\cellstack}[2][c]{\begin{tabular}[t]{@{}#1@{}}#2\end{tabular}}
\usepackage{orcidlink}
\usepackage[caption=false]{subfig}
\usepackage{commath}
\allowdisplaybreaks[1]

\usepackage{array}

\begin{document}

\title{Gyromagnetic ratios of slowly rotating charged Ellis--Bronnikov wormholes}

\author{Safiqul Islam \orcidlink{0000-0003-1373-4137}}
\email{Corresponding author: sislam@kfu.edu.sa}
\affiliation{Department of Mathematics and Statistics, College of Science,
King Faisal University, P.O. Box 400, Al Ahsa 31982, Saudi Arabia}

\author{Anirudh Pradhan
\orcidlink{0000-0002-1932-8431}}
\email{pradhan.anirudh@gmail.com}
\affiliation{Centre for Cosmology, Astrophysics and Space Science,
GLA University, Mathura-281\,406, Uttar Pradesh, India}

\author{Francisco Tello-Ortiz \orcidlink{0000-0002-7104-5746}}
\email{francisco.tello@ufrontera.cl}\affiliation{Departamento de Ciencias Físicas, Universidad de La Frontera,
Casilla 54-D, 4811186 Temuco, Chile.}

\author{Safyan Mukhtar \orcidlink{0000-0001-5415-6762}}
\email{smahmad@kfu.edu.sa}
\affiliation{Department of Mathematics and Statistics, College of Science,
King Faisal University, P.O. Box 400, Al Ahsa 31982, Saudi Arabia}

\begin{abstract}
Charged Ellis--Bronnikov wormholes, supported by a phantom scalar field in
Einstein--Maxwell theory, connect two asymptotically flat regions, so a
rotating member carries a magnetic dipole moment and a gyromagnetic ratio at
each mouth. We construct the slowly rotating family in closed form on all
three background branches and show that, within the stationary axial dipole
sector and with no applied magnetic field, the first-order solution is unique.
The two gyromagnetic ratios obey an exact deficit relation set by the static
geometry alone. It gives $g_\pm<2$ at both mouths, bounds the deficit by
two-thirds of the squared ratio of scalar to electric charge, and, through the
positivity of the reduced magnetic moment, gives each ratio the sign of the
mass of its own mouth; the two ratios coincide only for reflection-symmetric
throats, so their difference measures the asymmetry of the wormhole. Rotation
drags the two ends relative to each other in a definite sense, with a dragging
profile that decreases monotonically across the throat, and the
electromagnetic field carries a strictly positive share of the angular
momentum on every finite sphere. Toward the extremal edge at the positive
mouth the deficit closes cubically on every trajectory studied, because the
redshift factor at the throat vanishes quadratically; the coefficient depends
on the path. When the charge is switched off, the frame dragging reproduces
known neutral results, whereas on the massive neutral branch the gyromagnetic
ratios stay away from the Kerr--Newman value.
\end{abstract}

\maketitle

\section{Introduction}
\label{sec:intro}

A rotating, electrically charged body carries a magnetic dipole moment $\mu$
as well as an angular momentum $J$. Measured in units fixed by its mass $M$
and charge $Q$, their ratio defines the gyromagnetic ratio
$g=2\mu M/(QJ)$. A classical body with proportional mass and charge densities
has $g=1$, whereas the Kerr--Newman black hole has
$g=2$~\cite{newmanMetricRotatingCharged1965,carterGlobalStructureKerr1968},
the value of the Dirac electron. The value $2$ is not tied to that particular
solution: it holds for a slightly charged, stationary, \rev{axisymmetric} black
hole~\cite{waldBlackHoleUniform1974} and for \rev{the electrovac solutions
generated by Ernst's prescription~\cite{reinaGyromagneticRatioEinsteinMaxwell1975}}, and a charged loop
around a black hole interpolates between the classical value far from the hole
and $2$ near the horizon~\cite{garfinkleGyromagneticRatioBlack1990}. Additional
fields change the picture. \rev{Slowly rotating dilaton black holes have $g<2$ at nonzero dilaton
coupling~\cite{horneRotatingDilatonBlack1992}}, and in Einstein--Maxwell-scalar
theories the ratio depends on the couplings~\cite{qiuSlowlyRotatingBlack2021}.
The gyromagnetic ratio is therefore a compact measure of how matter and
geometry together shape the magnetic response of a rotating object.

Wormholes offer a geometry in which this response can be examined without a
horizon. The idea goes back to the Einstein--Rosen
bridge~\cite{Einstein:1935particle}; traversable wormholes became a
systematic subject with the work of Morris and
Thorne~\cite{Morris:1988wormholesa} and of Morris, Thorne and
Yurtsever~\cite{Morris:1988tu}, reviewed in
Ref.~\cite{Visser:1995cc}. The simplest example is the
Ellis--Bronnikov wormhole~\cite{Ellis:1973yv,Bronnikov:1973fh}, which joins two
asymptotically flat regions through a throat held open by a phantom scalar
field~\cite{bronnikovScalarFieldsSources2018}. \rev{The geometry has been
probed with test scalar fields~\cite{Blazquez-Salcedo:2022kaw}, and its shadow,
weak lensing and accretion images have been
computed~\cite{Alloqulov:2024olb,Ishkaeva:2023xny}, including images seen from
either side of the throat~\cite{Cui:2026cfx}; generalized Ellis--Bronnikov
wormholes have also been studied in $f(R)$ gravity~\cite{Sokoliuk:2022xcf}.}

Static Ellis--Bronnikov
wormholes are unstable~\cite{shinkaiFateFirstTraversible2002,gonzalezInstabilityWormholesSupported2009,bronnikovInstabilitiesWormholesRegular2012}.
\rev{Slow-rotation perturbation theory indicates that rotation weakens the radial
instability and may remove
it~\cite{azadAreSlowlyRotating2024,azadRadialPerturbationsEllisBronnikov2024};
nonlinear evolutions from axisymmetric initial data find that rotation slows
the collapse and, at the highest spins sampled, detect no growth of the throat
over the evolved times~\cite{shirokovRotationSlowsInstability2026}.}
Rotating Ellis--Bronnikov wormholes are known both
nonperturbatively~\cite{kleihausRotatingEllisWormholes2014,chewGeometrySpinningEllis2016}
and in slow
rotation~\cite{kasharginSlowlyRotatingWormholes2008,kasharginSlowlyRotatingScalar2008,azadQuadrupolePerturbationsSlowly2025}.
In Einstein--Maxwell theory with a minimally coupled phantom scalar the
wormholes can also carry electric
charge~\cite{gonzalezInstabilityChargedWormholes2009,makitaChargedMultisheetWormhole2026}.
This charged family has three classes of backgrounds, and its radial and
nonradial perturbations have been
analyzed~\cite{gonzalezInstabilityChargedWormholes2009,blazquez-salcedoRadialPerturbationsCharged2026,blazquez-salcedoNonradialPerturbationsStatic2026};
related instabilities appear in a theory with a scalar--Maxwell
coupling~\cite{Wu:2025wlz}.

Rotating, electrically charged wormholes have been studied along several
lines. Stationary generalizations of the Ellis--Bronnikov wormhole are
known~\cite{matosRotatingScalarField2006,volkovStationaryGeneralizationsBronnikovEllis2021},
as are rotating and magnetized phantom wormholes with a nonzero scalar--Maxwell
coupling~\cite{matosClassEinsteinMaxwell2010}. Cl\'ement computed a
gyromagnetic ratio for a ring wormhole with a single flat
end~\cite{clementSelfgravitatingCosmicRings1999}. Rotating
traversable wormholes in Einstein--Maxwell theory have been obtained from
sources with Misner--Dirac strings~\cite{clementRotatingTraversableWormholes2023},
and rotating charged wormholes with two ends have been found numerically in
Einstein--Dirac--Maxwell
theory~\cite{dzhunushalievRotatingWormholesEinsteinDiracMaxwell2026}; the
latter define masses, charges, magnetic moments and angular momenta at both
ends but do not report a gyromagnetic ratio. \rev{In a subsequent study the
same authors evaluate the gyromagnetic ratios at both ends of these numerical
solutions~\cite{Dzhunushaliev:2025ngw}. What has not been obtained is the
magnetic response of the regular charged Ellis--Bronnikov family, supported by
a minimally coupled phantom scalar, to slow rotation. For this family the two
gyromagnetic ratios can be found in closed form, and the question is how they
are fixed by the geometry and how they compare.}

In this paper we answer that question at first order in the rotation.
Within the stationary axial dipole sector and with no applied magnetic
field, the slowly rotating solution is unique and is obtained in closed form
on all three background branches, with the asymptotic quantities defined
separately at each mouth. The central result is an exact relation for the
gyromagnetic ratios,
\begin{equation*}
2-g_s=\frac{2D^2\,\mathcal B}{G_s\bigl(r_0^2\,w+3Q^2\mathcal B\bigr)},\qquad s=\pm ,
\end{equation*}
where $D$ is the scalar constant, $r_0$ the radial scale, $G_s$ the
normalization at the end $s$, and $w$ and $\mathcal B>0$ are background
quantities defined in Sec.~\ref{sec:deficit}. It shows that $g_s<2$ at both
mouths, a bound with content at a positive-mass mouth since a negative-mass
mouth has $g_s<0$, and, because the reduced moment $\mu_s/(Q_sJ_s)$ is positive, that
$g_s$ has the sign of the mass $M_s$ of its mouth; the two ratios differ
unless the background is reflection symmetric. Written in terms of the charges, the
relation bounds the deficit at the positive mouth by $2D^2/(3Q^2)$. The rotation also drags the two ends relative to each other: \rev{the negative end rotates in the sense of the
angular momentum $J_+$ measured at the positive mouth, the dragging rate
normalized to its negative-end value decreases monotonically from $1$ to $0$
across the wormhole, and the Komar angular momentum $J_{\rm K}$ satisfies
$0<J_{\rm K}/J_+<1$ on every finite-radius sphere, for nonzero rotation.} Toward the edge of the parameter domain at the positive mouth the
deficit closes cubically: along specified paths at fixed $D/r_0>0$,
$2-g_+\simeq\tfrac{16}{27}\epsilon^3$, where $\epsilon=1-Q/R_{\rm th}$ measures
the distance from extremality, while the symmetric path gives
$\tfrac{\pi^2}{3}\epsilon^3$. Both coefficients follow from one exact
form of the deficit relation, in which the deficit is proportional to the
redshift factor at the throat; near the edge this factor vanishes as
$\epsilon^2$. When the
charge is switched off, the frame dragging reproduces the known neutral results
of Kashargin and Sushkov~\cite{kasharginSlowlyRotatingWormholes2008}, but the
gyromagnetic ratios do not tend to $2$, in contrast with dilaton black holes.
All results are stationary and of first order in the rotation; we make no
claim about dynamical stability.

The paper is organized as follows. Section~\ref{sec:background} describes the
theory and the static backgrounds, and Sec.~\ref{sec:firstorder} constructs
the slowly rotating solution. Section~\ref{sec:results} derives the
gyromagnetic ratios and the frame dragging, and Sec.~\ref{sec:extremal} treats
the limiting behavior and the comparison with black holes.
Section~\ref{sec:conclusions} summarizes our conclusions. Appendix~\ref{app:branches} collects the
closed-form expressions, and Appendix~\ref{app:proofs} the proofs, the Komar
integral, the limiting expansions and the numerical checks.

\section{Theory and static backgrounds}
\label{sec:background}

The wormholes studied here are charged generalizations of the
Ellis--Bronnikov solution~\cite{Ellis:1973yv,Bronnikov:1973fh}.

\subsection{Theory and conventions}
\label{sec:theory}

We consider Einstein--Maxwell theory with a minimally coupled phantom scalar
field, in signature $(-,+,+,+)$ and units $G_{\rm N}=c=1$,
\begin{equation}
I=\frac{1}{16\pi}\int d^4x\,\sqrt{-g}\,
\bigl[R-F_{\mu\nu}F^{\mu\nu}+2\nabla_\mu\phi\nabla^\mu\phi\bigr].
\label{eq:action}
\end{equation}
The field equations are the Einstein equations sourced by the Maxwell and
scalar stress tensors, together with $\nabla_\mu F^{\mu\nu}=0$ and
$\Box\phi=0$. The sign of the scalar kinetic term makes $\phi$ a phantom
field; it supplies the violation of the null energy condition that keeps the
throat open. With minimal coupling, a regular static throat of this kind
requires a scalar with negative kinetic energy, as reviewed in
Ref.~\cite{bronnikovScalarFieldsSources2018}.

\subsection{Static backgrounds}
\label{sec:backgrounds}

The static, spherically symmetric configurations can be written as
\begin{equation}
\begin{gathered}
ds_0^2=-G\,dt^2+G^{-1}\bigl[dr^2+S\,d\Omega^2\bigr],\\
\mathcal A_0=V\,dt,\qquad \phi=\phi(r),
\end{gathered}
\label{eq:static}
\end{equation}
with $S=r^2+r_0^2$, $r_0>0$ and $r\in\mathbb R$. The areal radius is
$R=\sqrt{S/G}$. The parameter $r_0$ sets the radial scale, whereas the throat
radius $R_{\rm th}$ is the minimum areal radius; the two need not coincide.
With the charge normalized by $\sqrt{-g}\,F^{rt}=Q\sin\theta$, the Maxwell and
scalar equations integrate once,
\begin{equation*}
V'=-\frac{QG}{S},\qquad \phi'=\frac{D}{S},
\end{equation*}
where $Q$ is the electric charge and $D$ the scalar integration constant.
Charged backgrounds are presented with $Q>0$. Reversing the sign of $Q$
reverses $V$ and, at first order in the rotation, also $p$ and $\mu_s$, leaving
$g_s$ unchanged.

The remaining Einstein equations are simplest for $W=G^{-1/2}$ as a function
of $u=\pi/2-\arctan(r/r_0)\in(0,\pi)$, which places the positive end at $u=0$
and the negative end at $u=\pi$. With $k=D^2-r_0^2$, $a=k/r_0^2$ and
$q=Q/r_0$, they become
\begin{equation}
W_{uu}=aW,\qquad W_u^2-aW^2=q^2,\qquad W(0)=1,
\label{eq:Weq}
\end{equation}
where $W(0)=1$ normalizes the time $t$ to the positive end. The solution is
\begin{equation}
W=C_a(u)+m\,S_a(u),\qquad m\equiv\frac{M_+}{r_0},\qquad m^2=q^2+a,
\label{eq:Wunified}
\end{equation}
where $C_a$ and $S_a$ solve $d^2f/du^2=af$ with $C_a(0)=1$, $C_a'(0)=0$ and
$S_a(0)=0$, $S_a'(0)=1$:
\begin{equation}
(C_a,S_a)=
\begin{cases}
(\cos\eta u,\ \eta^{-1}\sin\eta u), & a=-\eta^2,\\
(1,\ u), & a=0,\\
(\cosh\lambda u,\ \lambda^{-1}\sinh\lambda u), & a=\lambda^2 .
\end{cases}
\label{eq:CS}
\end{equation}
They satisfy $C_a^2-aS_a^2=1$, and $(D/r_0)^2=1+a$. The mass $M_+$ of the
positive end is defined in Sec.~\ref{sec:ends}; the sign of $m=\pm\sqrt{q^2+a}$
is that of $M_+$.

The sign of $k$ divides the backgrounds into three classes, listed in
Table~\ref{tab:backgrounds}. On the critical branch, $k=0$, $W=1+qu$, and
integration of the Maxwell equation gives $V=V_0-W^{-1}$, where $V_0$ is a
gauge constant. The hyperbolic, or subcritical, branch has
$k=\lambda^2r_0^2>0$ and a hyperbolic $W$. The trigonometric, or
supercritical, branch has $k=-\eta^2r_0^2<0$; with $\alpha=\pi\eta/2$ and
$\theta_+=\alpha+\delta$, its unified form reads
$W=\cos\eta u+\tan\theta_+\sin\eta u$, and its member $\delta=0$ is symmetric
under $r\to-r$. \rev{At fixed $r_0$, the limit $Q\to0$ recovers the massless
Ellis wormhole on the critical branch and the massive Ellis--Bronnikov wormhole
on the hyperbolic branch at fixed $\lambda>0$. The trigonometric branch has no
neutral member at fixed $\eta>0$, since $Q\ge\eta r_0$, but approaches the
massless Ellis wormhole as $\eta\to0$ at fixed admissible $\delta$.} These three classes appear in the work of Gonz\'alez, Guzm\'an
and Sarbach~\cite{gonzalezInstabilityChargedWormholes2009}. More recently,
Makita et al.\ used a Harrison transformation to construct related charged
wormholes with any even number of asymptotically flat
regions~\cite{makitaChargedMultisheetWormhole2026}.

\begin{table*}
\caption{\label{tab:backgrounds}Regular static charged Ellis--Bronnikov backgrounds with two asymptotically flat ends, $ds^2=-G\,dt^2+G^{-1}[dr^2+(r^2+r_0^2)\,d\Omega^2]$, classified by the sign of $k=D^2-r_0^2$. Here $y=\arctan(r/r_0)$, $u=\tfrac{\pi}{2}-y\in(0,\pi)$, $q=Q/r_0$, $G_+=1$ ($t$ is the $+$-end time), $Q_\pm=\pm Q$, and $M_s=-s\,r_0\,W_y(s\tfrac{\pi}{2})$ with $W=G^{-1/2}$; $W$ obeys $W_{yy}=aW$ and $W_y^2-aW^2=q^2$ with $a=k/r_0^2$. The parameter domains are the regularity conditions. The critical and hyperbolic columns are written in the orientation $M_+>0$; the opposite orientation is the same family with the ends exchanged ($r\to-r$) and $G_+=1$ restored. The trigonometric column contains both orientations. The throat is at $r=r_{\rm th}$, $u=u_{\rm th}$, with areal radius $R_{\rm th}=\min_r\sqrt{r^2+r_0^2}\,W$.}
\begin{ruledtabular}
\begin{tabular}{lccc}
 & \cellstack{Critical,\\ $k=0$} & \cellstack{Hyperbolic (subcritical),\\ $k=\lambda^2r_0^2>0$} & \cellstack{Trigonometric (supercritical),\\ $k=-\eta^2r_0^2<0$\footnotemark[1]}\\
\colrule
Parameters & $q>0$ & $\lambda>0$, $q>0$ & \cellstack{$0<\eta<1$, $|\delta|<\tfrac{\pi}{2}(1-\eta)$\footnotemark[2];\\ $\alpha=\tfrac{\pi\eta}{2}$, $c_\pm=\cos(\alpha\pm\delta)>0$}\\[2pt]
$Q$ & $qr_0$ & $qr_0$ & $\eta r_0/c_+$\\[2pt]
$D$ & $r_0$ & $r_0\sqrt{1+\lambda^2}$ & $r_0\sqrt{1-\eta^2}$\\[2pt]
$W=G^{-1/2}$ & $1+q\big(\tfrac{\pi}{2}-y\big)$ & \cellstack{$\dfrac{q}{\lambda}\sinh\!\Big[\lambda\Big(\dfrac{\pi}{2}-y\Big)+\zeta\Big]$,\\ $\sinh\zeta=\lambda/q$} & $\dfrac{\cos(\eta y+\delta)}{c_+}$\\ \noalign{\vskip5pt}
$\mathcal{B}$ [Eq.~\eqref{eq:Bdef}] & $1$ & $\dfrac{\cosh\pi\lambda}{1+4\lambda^2}$ & $\dfrac{\cos\pi\eta}{1-4\eta^2}$ \ ($\tfrac{\pi}{4}$ at $\eta=\tfrac12$)\\ \noalign{\vskip3pt}
$G_-$ & $(1+\pi q)^{-2}$ & $\sinh^2\!\zeta/\sinh^2(\zeta+\pi\lambda)$ & $c_+^2/c_-^2$\\[2pt]
$M_+$ & $Q$ & $r_0\sqrt{q^2+\lambda^2}$ & $Q\sin(\alpha+\delta)$\\[2pt]
$M_-$ & $-Q$ & $-Q\cosh(\zeta+\pi\lambda)$ & $Q\sin(\alpha-\delta)$\\[2pt]
$Q$ vs $M_+$ & $Q=M_+$ & $Q<M_+$ & $|M_+|<Q$\\[2pt]
$r_{\rm th}/r_0$ & $q/(1+qu_{\rm th})>0$ & $\lambda\coth(\lambda u_{\rm th}+\zeta)>0$ & \cellstack{$\eta\tan(\eta y_{\rm th}+\delta)$,\\ sign of $\delta$}\\[2pt]
$R_{\rm th}/r_0$\footnotemark[3] & $>1$ & $>1$ & \cellstack{$\big(\eta/\sin\pi\eta,\,\infty\big)$ at fixed $\eta$,\\ increasing in $\delta$}\\
\end{tabular}
\end{ruledtabular}
\footnotetext[1]{Symmetric member $\delta=0$: throat at $r=0$, $R_{\rm th}=r_0\sec\alpha$, $Q/R_{\rm th}=\eta$, $M_-=M_+=Q\sin\alpha>0$.}
\footnotetext[2]{Band edges: as $\delta\to[\tfrac{\pi}{2}(1-\eta)]^-$, $Q,R_{\rm th}\to\infty$ with $Q/R_{\rm th}\to1$ (the positive-mouth extremal edge of Table~\ref{tab:limits}); as $\delta\to[-\tfrac{\pi}{2}(1-\eta)]^+$, $G_-\to\infty$ and $R_{\rm th}\to Q=\eta r_0/\sin\pi\eta$ (the negative-mouth extremal limit).}
\footnotetext[3]{The throat is unique, and Eq.~\eqref{eq:throatid} gives $Q<R_{\rm th}$. The parameter $r_0$ sets the radial scale, whereas $R_{\rm th}$ is the minimum areal radius; $R_{\rm th}/r_0$ can be below, equal to or above unity on the trigonometric branch (Sec.~\ref{sec:throat}).}
\end{table*}

A background is regular, with two asymptotically flat ends, when
$0<W<\infty$ on the closed interval $0\le u\le\pi$. An interior zero of $W$
would make the areal radius vanish at finite $r$, a curvature singularity;
the parameter domains of Table~\ref{tab:backgrounds} are exactly the
regularity conditions. Because $G>0$ everywhere, none of these backgrounds
has a horizon. The critical and hyperbolic columns are written with $M_+>0$;
the opposite orientation is the same family with the ends exchanged,
$r\to-r$, and $G_+=1$ restored. The trigonometric family contains both
orientations. The radial stability of the family was analyzed by Gonz\'alez,
Guzm\'an and Sarbach~\cite{gonzalezInstabilityChargedWormholes2009}, who
proved instability for the critical and subcritical classes and gave
numerical evidence for it in the supercritical class; radial and nonradial
perturbations have since been studied in more
detail~\cite{blazquez-salcedoRadialPerturbationsCharged2026,blazquez-salcedoNonradialPerturbationsStatic2026}.
Here we study the stationary response of these backgrounds to first order in
rotation.

\subsection{Asymptotic ends}
\label{sec:ends}

Each end is asymptotically flat, and its mass and charge are read off in its
own normalization. We label the ends by $s=\pm$, with $sr\to+\infty$ at end
$s$, and write $G_s$ for the limit of $G$ there, so that $G_+=1$. With
$t_s=\sqrt{G_s}\,t$ the proper time at that infinity,
\begin{equation}
g_{t_st_s}=-1+\frac{2M_s}{R}+\dots,\qquad
A_{t_s}={\rm const}+\frac{Q_s}{R}+\dots .
\label{eq:MQdefs}
\end{equation}
Section~\ref{sec:firstorder} adds a uniformly rotating azimuth at each end
[Eq.~\eqref{eq:frames}]. In terms of
$y=\arctan(r/r_0)$,
\begin{equation*}
Q_s=sQ,\qquad M_s=-s\,r_0\,W_y\bigl(s\tfrac{\pi}{2}\bigr),
\end{equation*}
so that $M_+=r_0m$, $M_-=-r_0W_u(\pi)$ and $G_-^{-1/2}=W(\pi)$. The charges
are opposite because the two ends face in opposite directions. The scalar
field approaches its asymptotic values as
$\phi=\phi_s-sD/(\sqrt{G_s}\,R)+O(R^{-2})$, so the coefficient of $1/R$
also depends on the end.

The masses of the two ends need not agree, and one of them can be negative.
On the critical and hyperbolic branches $M_+>0>M_-$. On the trigonometric
branch $M_-<0$ for $\delta>\alpha$ and $M_+<0$ for $\delta<-\alpha$; either
case requires $\eta<\tfrac12$, and the two masses are never both negative. At
the positive end, $m^2=q^2+a$ orders the charge and the mass: $Q=M_+$ on the
critical branch, $Q<M_+$ on the hyperbolic branch and $|M_+|<Q$ on the
trigonometric branch. These relations organize the mass--charge plane of
Sec.~\ref{sec:gmaps}.

\subsection{Throat geometry}
\label{sec:throat}

Since $S=r_0^2/\sin^2u$, the areal radius is $R=r_0W/\sin u$, and its minimum
satisfies
\begin{equation}
W_u\sin u-W\cos u=0 .
\label{eq:throatcond}
\end{equation}
Every background has exactly one throat. With $F=\cot u-W_u/W$, the identity
$F_u+(r/r_0+W_u/W)F=-(1+a)=-(D/r_0)^2$ holds on every branch, so $F$
decreases through each of its zeros; since $F\to+\infty$ as $u\to0$ and
$F\to-\infty$ as $u\to\pi$, it has exactly one. At the throat
$W_u=W\cot u$, and the constraint in Eq.~\eqref{eq:Weq} gives the throat
identity
\begin{equation}
1-\frac{Q^2}{R_{\rm th}^2}=(1+a)\sin^2u_{\rm th}=\frac{D^2}{r_0^2+r_{\rm th}^2}.
\label{eq:throatid}
\end{equation}
Hence $0<Q/R_{\rm th}<1$ for every charged background. Section~\ref{sec:extremal}
examines specified limiting trajectories along which $Q/R_{\rm th}\to1$.

The throat radius and $r_0$ are related differently on the three branches.
For the charged members of the critical and hyperbolic branches, $W>1$ at
every finite $r$, so $R_{\rm th}>r_0$; the neutral massless endpoint, $W\equiv1$,
has $R_{\rm th}=r_0$. On the trigonometric branch $W$ can fall below unity. At
fixed $\eta$, $d\ln(R_{\rm th}/r_0)/d\delta=\tan(\alpha+\delta)
-\tan(\eta y_{\rm th}+\delta)>0$, so $R_{\rm th}/r_0$ increases from
$\eta/\sin\pi\eta$ to infinity across the band. The infimum over the whole
branch, $1/\pi$, is not attained, and throats smaller than $r_0$ occur only
for $\eta<\eta_*$, where $\sin\pi\eta_*=\eta_*$ and $\eta_*\simeq0.73648445$.

\begin{figure*}[!tp]
\centering
\includegraphics[width=\textwidth]{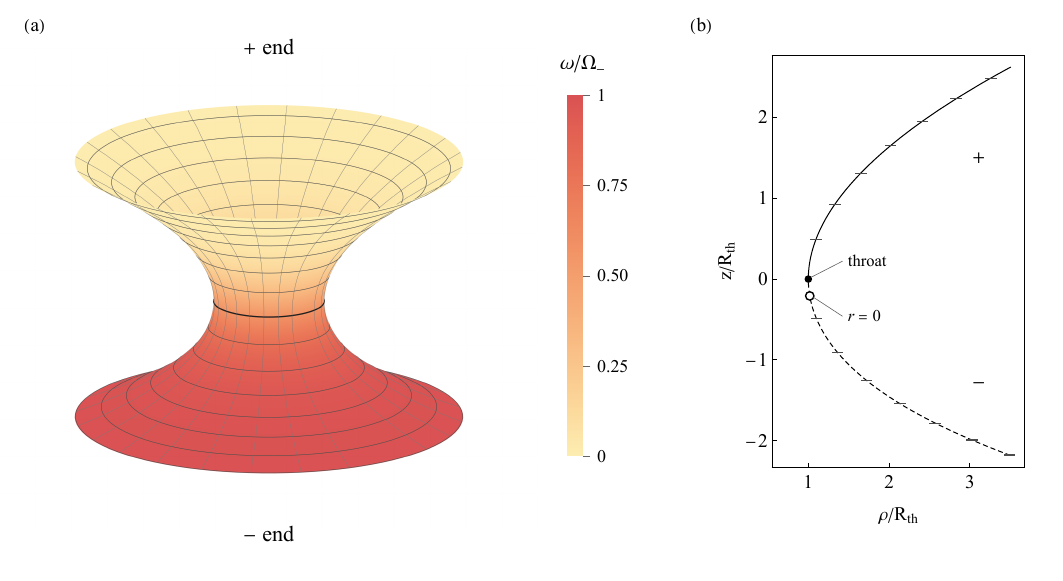}
\caption{Equatorial embedding of the trigonometric background with
$(\eta,\delta)=(2/5,2/5)$, for which $M_+>M_->0$. (a) Surface of revolution,
colored by the frame-dragging profile $\omega/\Omega_-$, which falls
monotonically from 1 at the negative end to 0 at the positive end, where the
frame does not rotate; the rings are separated by a
proper distance $0.5R_{\rm th}$. (b) Meridional profile, in units of $R_{\rm th}$
and with equal axis scales, with the positive sheet solid and the negative sheet dashed; the filled circle
marks the throat and the open circle $r=0$. The embedding represents the
static spatial geometry, which is unchanged at first order in rotation. Both
sheets are cut at $R=3.5R_{\rm th}$.}
\label{fig:embedding}
\end{figure*}

Figure~\ref{fig:embedding} shows the equatorial embedding of an asymmetric
trigonometric background. The two sheets meet at a single throat, which lies
at $r>0$ because $\delta>0$, and the two ends have different masses. The
coloring anticipates the frame-dragging profile constructed in
Sec.~\ref{sec:firstorder} and discussed in Sec.~\ref{sec:framedrag}.

\section{Slowly rotating solution}
\label{sec:firstorder}

We now set the static wormholes of Sec.~\ref{sec:background} into slow
rotation. Rotation drags inertial frames and, because the configuration is
charged, also produces a magnetic field. At first order the two effects are
coupled through a pair of ordinary differential equations in $r$, which can be
solved in closed form on every background class.

\subsection{Slow-rotation equations}
\label{sec:sloweqs}

We perturb the static background by a small rotation parameter
$\varepsilon_{\rm rot}$ and write, in the slow-rotation form of
Hartle~\cite{hartleSlowlyRotatingRelativistic1967},
\begin{equation}
\begin{aligned}
ds^2&=ds_0^2-2\varepsilon_{\rm rot}\,H\,\omega(r)\sin^2\theta\,dt\,d\varphi,\\
\mathcal A&=V\,dt+\varepsilon_{\rm rot}\,p(r)\sin^2\theta\,d\varphi,
\end{aligned}
\label{eq:ansatz}
\end{equation}
where $ds_0^2$ and $V$ are the background metric and electric potential, and
\begin{equation}
S=r^2+r_0^2,\qquad H=\frac{S}{G}=R^2 ,
\label{eq:SHdefs}
\end{equation}
with $R$ the areal radius. Throughout, ``first order'' refers to this
expansion in $\varepsilon_{\rm rot}$. The static slice, the throat radius
$R_{\rm th}$ and the masses $M_\pm$ keep their background values; the
deformation of the throat by rotation first appears at second order. The
perturbation belongs to the stationary, axisymmetric axial dipole sector. In
this sector the scalar field is unperturbed, and the component $h_{r\varphi}$
is removed by the coordinate change $\varphi\to\varphi+\varepsilon_{\rm rot}\Lambda(r)$,
which shifts it by $(S/G)\Lambda'\sin^2\theta$. Polar perturbations are even
under $(t,\varphi)\to(-t,-\varphi)$ and are not excited at first order, and axial
modes with $\ell\ge2$ carry no source; the uniqueness statement below refers to
the $\ell=1$ axial sector.

For the neutral Ellis--Bronnikov wormhole this sector was first studied by
Kashargin and Sushkov~\cite{kasharginSlowlyRotatingWormholes2008}, who later
extended the analysis to second order~\cite{kasharginSlowlyRotatingScalar2008}.
Subsequent studies have examined the radial perturbations of slowly rotating
Ellis--Bronnikov wormholes at second order and their quadrupole
sector~\cite{azadRadialPerturbationsEllisBronnikov2024,azadQuadrupolePerturbationsSlowly2025},
and rapidly rotating neutral solutions are known
numerically~\cite{kleihausRotatingEllisWormholes2014,chewGeometrySpinningEllis2016}.
The charge adds one new ingredient: the magnetic potential $p$, which couples
to the frame dragging already at first order.

Only the $t\varphi$ component of the Einstein equations and the $\varphi$
component of the Maxwell equations are nontrivial at this order. The phantom
scalar does not contribute to $T^t{}_\varphi$ and enters only through $G$.
With a prime denoting $d/dr$,
\begin{equation}
(H^2\omega')'=-4Q\,p',\qquad
(Gp')'+Q\,\omega'-\frac{2G}{S}\,p=0 .
\label{eq:firstorder}
\end{equation}
The first equation integrates once,
\begin{equation}
H^2\omega'+4Qp=C ,
\label{eq:Cdef}
\end{equation}
with a constant $C$ that will be related to the angular momentum below.
From here onward we absorb $\varepsilon_{\rm rot}$ into $p$, $\omega$, $C$
and their asymptotic coefficients, retaining the same symbols for the
physical quantities at first order. The ratios used in
Sec.~\ref{sec:results} are unchanged.

\subsection{Solution and asymptotic quantities}
\label{sec:solution}

Eliminating $\omega'$ with Eq.~\eqref{eq:Cdef} leaves a single equation for
the magnetic potential,
\begin{equation}
\begin{aligned}
(Gp')'-Up&=-\frac{CQG^2}{S^2},\\
U&=\frac{2G}{S}+\frac{4Q^2G^2}{S^2}>0 .
\end{aligned}
\label{eq:master}
\end{equation}
We look for solutions with
\begin{equation}
p(\pm\infty)=0,\qquad \omega(+\infty)=0,\qquad \omega(-\infty)=\Omega_- .
\label{eq:bc}
\end{equation}
The conditions on $p$ and the normalization $\omega(+\infty)=0$ are imposed;
the last relation defines $\Omega_-$, which the solution then determines. At large $|r|$ the homogeneous solutions of
Eq.~\eqref{eq:master} behave as $|r|^2$ or $1/|r|$. The growing one is a
uniform magnetic field, so decay of $p$ at both ends means that no external
field is applied. Because only $\omega'$ enters the equations, $\omega$ is
defined up to a constant, a rigid rotation of the coordinates; the condition
$\omega(+\infty)=0$ fixes it, and $\Omega_-$ is then the rotation of the
negative end relative to the positive one, measured in the time of the
positive end. Of the four integration constants, the three conditions leave
one free: the rotation amplitude $C$.

Let $p_+$ and $p_-$ be solutions of the
homogeneous equation ($C=0$) that decay at the positive and negative ends,
respectively. They are independent (Sec.~\ref{sec:unique}) and define the
Green function of Eq.~\eqref{eq:master}, which gives the unique decaying $p$.
The frame-dragging function then follows from the first integral,
$\omega(r)=-\int_r^{\infty}(C-4Qp)\,H^{-2}\,dr'$. On each of the three
background classes the homogeneous solutions are elementary, so $p$ and
$\omega$ are obtained in closed form. On the critical branch they follow from
Israel--Wilson--Perj\'es-type modes~\cite{IsraelWilson:1972,Perjes:1971gv}; on the
trigonometric branch, from an elementary ansatz. The hyperbolic expressions
follow by the continuation $\eta=i\lambda$, $\delta=\pi/2-i\beta$, with
$\beta=\zeta+\pi\lambda/2$. At $\eta=\tfrac12$, the chosen trigonometric
basis degenerates. The regular expressions for $p$ and $\omega$ at this
parameter value are given in Appendix~\ref{app:branches}. We use these
branch-specific closed forms; the
equations and the Green-function construction apply to all three branches.

The angular momentum and magnetic moment are read off in an asymptotic frame
at each end $s=\pm$,
\begin{equation}
t_s=\sqrt{G_s}\,t,\qquad \varphi_s=\varphi-\omega_s t,\qquad
\omega_+=0,\quad \omega_-=\Omega_- ,
\label{eq:frames}
\end{equation}
where $G_s$ is the limit of $G$ at that end. The time is normalized to
proper time at infinity, the constant rotation of the frame is removed, and
the azimuth keeps the same orientation at both ends. In these frames we write
\begin{equation}
\begin{gathered}
g_{t_st_s}=-1+\frac{2M_s}{R},\qquad A_{t_s}={\rm const}+\frac{Q_s}{R},\\
g_{t_s\varphi_s}=-\frac{2J_s}{R}\sin^2\theta,\qquad
A_{\varphi_s}=-\frac{\mu_s}{R}\sin^2\theta ,
\end{gathered}
\end{equation}
up to higher powers of $1/R$, with the masses $M_s$ of
Sec.~\ref{sec:background}. The first integral fixes the tail of $\omega$, and
the decaying solution defines the dipole coefficient $b_s$:
\begin{equation}
p=\frac{b_s}{|r|}+O\bigl(|r|^{-2}\bigr),\qquad
\omega=\omega_s-\frac{sCG_s^2}{3|r|^3}+O\bigl(|r|^{-4}\bigr).
\label{eq:tails}
\end{equation}
Since $R\simeq|r|/\sqrt{G_s}$, this gives
\begin{equation}
J_s=-\frac{sC}{6},\qquad \mu_s=-\frac{b_s}{\sqrt{G_s}},\qquad Q_s=sQ .
\label{eq:asym}
\end{equation}
Hence $C=-6J_+$ and $J_-=-J_+$, just as $Q_-=-Q_+$: the two ends face in
opposite directions, while the azimuth keeps one orientation. Reversing the
azimuth at one end would flip $J_s$ and $\mu_s$ together and leave $g_s$
unchanged. The factor $G_s^{-1/2}$ in $\mu_s$ converts the magnetic
tail to the normalization of the end $s$; it is unity at the positive end.
For $Q\ne0$ and $C\ne0$ we define the gyromagnetic ratio and the reduced
moment at each end,
\begin{equation*}
g_s=\frac{2\mu_sM_s}{Q_sJ_s},\qquad \gamma_s=\frac{\mu_s}{Q_sJ_s} .
\end{equation*}

\subsection{Uniqueness and positivity}
\label{sec:unique}

For a regular background and fixed $C$, Eqs.~\eqref{eq:firstorder} with the
conditions~\eqref{eq:bc} have exactly one solution $(p,\omega)$; this holds
also for $Q=0$ or $C=0$. The difference of two solutions satisfies the
homogeneous master equation and decays at both ends. Multiplying by the
difference and integrating over $r$, the positivity of $G$ and $U$ forces it
to vanish. The same argument shows that $p_+$ and $p_-$ cannot be
proportional, so their Wronskian does not vanish and the Green function
exists. At a fixed background the solution is linear in $CQ$,
$p=CQ\,f(r)$, where $f$ depends only on the background. Consequently
$\mu_s\propto QJ_+$, and $g_s$ and $\gamma_s$ depend only on the background
parameters.

For a charged, rotating configuration the magnetic response has a definite
sign. When $Q\ne0$ and $C\ne0$, $f>0$ at every finite $r$, and its dipole
coefficient at each end is strictly positive. With Eq.~\eqref{eq:asym} this
gives
\begin{equation}
\gamma_s=\frac{\mu_s}{Q_sJ_s}>0,\qquad
\operatorname{sign}g_s=\operatorname{sign}M_s .
\label{eq:gamma}
\end{equation}
The function $f$ obeys the master equation with the strictly negative source
$-G^2/S^2$ and vanishes at both ends, so it cannot have a nonpositive interior
minimum. If its $1/|r|$ coefficient vanished at one end, the equation would
give $f=-G_s/(4r^2)+O(|r|^{-3})<0$ there; the coefficient is therefore
positive. The proofs of uniqueness and positivity are given in
Appendix~\ref{app:proofs}. At $Q=0$ the magnetic potential vanishes
identically and $g_s$ and $\gamma_s$ are undefined, whereas $\omega$ remains
well defined; the limits of the ratios as $Q\to0$ are discussed in
Sec.~\ref{sec:neutral}.

The closed forms satisfy both first-order equations and their first
integral, and agree with an independent numerical solution of the
boundary-value problem; near extremality the deficit is evaluated from a
manifestly positive closed form (Appendix~\ref{app:checks}).

\section{Gyromagnetic ratios and frame dragging}
\label{sec:results}

The slow-rotation solution determines the magnetic dipole moment and angular
momentum at each asymptotic end, and their ratio need not be the same at the
two mouths. With the definitions of Sec.~\ref{sec:solution},
\begin{equation}
g_s=2M_s\gamma_s .
\label{eq:g-gamma}
\end{equation}
We retain $r_0$ explicitly and use $q=Q/r_0$ and $a=k/r_0^2$, with
$k=D^2-r_0^2$.

\subsection{The exact deficit relation}
\label{sec:deficit}

All three background classes obey the same relation at both mouths:
\begin{equation}
\begin{aligned}
2-g_s
&=\frac{2(D/r_0)^2\,\mathcal B}{G_s\bigl(w+3q^2\mathcal B\bigr)}\\
&=\frac{2D^2\,\mathcal B}{G_s\bigl(r_0^2w+3Q^2\mathcal B\bigr)},
\qquad s=\pm .
\end{aligned}
\label{eq:deficit}
\end{equation}
Here $G_+=1$ and $w=W(\pi)=G_-^{-1/2}$ is the negative-end value of
$W=C_a(u)+(M_+/r_0)S_a(u)$, defined in Sec.~\ref{sec:background}. The branch
factor is
\begin{equation}
\mathcal B=\frac{C_a(\pi)}{1+4a}
=\begin{cases}
\dfrac{\cos\pi\eta}{1-4\eta^2}, & a=-\eta^2,\\[6pt]
1, & a=0,\\[4pt]
\dfrac{\cosh\pi\lambda}{1+4\lambda^2}, & a=\lambda^2.
\end{cases}
\label{eq:Bdef}
\end{equation}
On the trigonometric branch the numerator and denominator change sign
together at $\eta=\tfrac12$, where the continuous value is $\pi/4$.
Consequently $\mathcal B>0$ on every branch. All factors in the denominator
of Eq.~\eqref{eq:deficit} are also positive for a regular background, and
therefore $g_s<2$ at both mouths. This is an upper bound on $g_s$, not on its
absolute value. Values below 2 are not unusual away from the Kerr--Newman
family: for a charged loop around a black hole, for example, $g$ rises from the
classical value 1 far from the hole to 2 near the
horizon~\cite{garfinkleGyromagneticRatioBlack1990}.

Equation~\eqref{eq:deficit} can also be written as
\begin{equation}
2-g_+=\frac{2}{3}\,\frac{D^2}{Q^2}\left[1+\frac{r_0^2w}{3Q^2\mathcal B}\right]^{-1},
\label{eq:bound}
\end{equation}
so the deficit at the positive mouth is bounded by $2D^2/(3Q^2)$ and approaches
this bound where $r_0^2w\ll Q^2\mathcal B$; at the negative mouth the bound
carries the factor $G_-^{-1}$. The scalar charge therefore sets the scale of the
deficit relative to the electric charge, but it does not act alone:
$\mathcal B$ and $w$ also depend on $D$ through $a$, and on the critical branch,
where $D=r_0$ is fixed, the variation of $g_+$ comes entirely from $q$.
Formally, setting $D=0$ in the same equations gives the slowly rotating
Kerr--Newman solution, for which Eq.~\eqref{eq:deficit} vanishes
(Appendix~\ref{app:komar}); its static background is Reissner--Nordstr\"om
with $Q>M_+$, a naked singularity outside the regular family.

The positivity of $\gamma_s$ established in Sec.~\ref{sec:firstorder}
[Eq.~\eqref{eq:gamma}] gives a second result:
$\operatorname{sign}g_s=\operatorname{sign}M_s$. A negative gyromagnetic
ratio accompanies a negative-mass mouth. For example, the critical solution
with $Q/r_0=1$ has $g_+=1.71995$ and $g_-=-2.80363$. Near the mass-zero
boundaries the first-order value of $g_s$ requires $J^2\ll|M_s|\,r_0^3$, since
the masses are corrected at second order in the
rotation~\cite{azadAreSlowlyRotating2024,azadRadialPerturbationsEllisBronnikov2024}.
The reduced moment $\gamma_s$, which does not change sign with the mass, is the
cleaner response quantity; its scale-free form $R_{\rm th}\gamma_s$ is shown in
Fig.~\ref{fig:asym}(b).

Equation~\eqref{eq:deficit} follows by extracting the dipole coefficients
of the $1/|r|$ tails of $p$ and combining them with the asymptotic masses,
charges and angular momenta. Appendix~\ref{app:branches} gives the
intermediate steps and the separate branch formulas. The approach to the
Kerr--Newman value $g=2$~\cite{newmanMetricRotatingCharged1965,carterGlobalStructureKerr1968} along specified
near-extremal trajectories is examined in Sec.~\ref{sec:extremal}.

\subsection{Gyromagnetic ratios over the mass--charge plane}
\label{sec:gmaps}

\begin{figure*}[!tp]
\centering
\includegraphics[width=\textwidth]{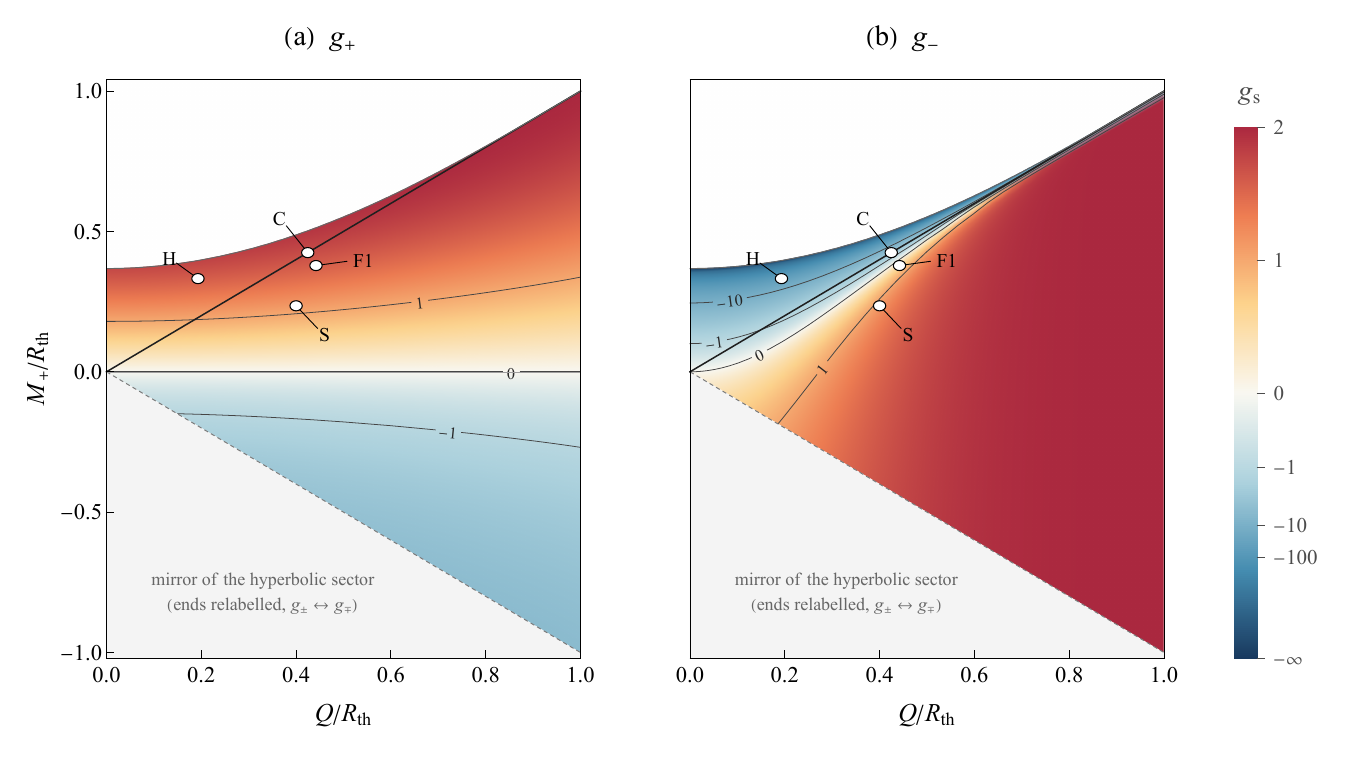}
\caption{Gyromagnetic ratios at the two mouths, (a) $g_+$ and (b) $g_-$, in
the plane $X=Q/R_{\rm th}$, $Y=M_+/R_{\rm th}$. The trigonometric family
occupies $-X<Y<X$; the critical family lies on $Y=X$; and the colored
hyperbolic sector lies between this line and the upper boundary. Gray denotes
regular hyperbolic configurations with the ends exchanged. Color is linear
for $g_s\ge0$ and uses inverse-hyperbolic-sine scaling for $g_s<0$; the bound
$g_s<2$ holds throughout the regular family. Thin contours indicate
$g_+=-1,0,1$ in (a) and $g_-=-10,-1,0,1$ in (b). The zero contours coincide
with the corresponding mass-zero boundaries; near them the first-order values
require $J^2\ll|M_s|\,r_0^3$. Markers \textsf{C}, \textsf{S}, \textsf{F1} and \textsf{H} identify
the configurations of Fig.~\ref{fig:profiles}; \textsf{F1} is also used in
Fig.~\ref{fig:embedding}.}
\label{fig:gmaps}
\end{figure*}

Figure~\ref{fig:gmaps} compares the two ratios using
$X=Q/R_{\rm th}$ and $Y=M_+/R_{\rm th}$. The areal throat radius supplies a
geometric length scale common to all three families. The throat identity
gives $0<X<1$ for charged regular solutions. The trigonometric family occupies
$-X<Y<X$, the positive-orientation critical family lies on $Y=X$, and the
positive-orientation hyperbolic family occupies $X<Y<Y_{\max}(X)$, where
\begin{equation}
Y_{\max}(X)=\cosh t-t\sinh t,\qquad t=\sqrt{1-X^2} .
\label{eq:Ymax}
\end{equation}
The upper boundary increases from $1/e$ at $X=0$ to $1$ as $X\to1$, since
$dY_{\max}/dX=X\cosh t>0$. At $X=0$, the value $1/e$ is approached by the
neutral hyperbolic family as $\lambda\to\infty$.

Exchanging the mouths gives the opposite critical orientation on $Y=-X$
and the mirror hyperbolic sector below it. The latter is shaded gray in the
figure; its gyromagnetic ratios follow by interchanging $g_+$ and $g_-$.
Appendix~\ref{app:branches} derives the chart and shows that each
configuration, up to the overall length scale, has a unique position in the
displayed sectors.

The two panels show how the mass signs affect the magnetic response.
In panel (a), $g_+$ changes sign across $Y=0$; in panel (b), the corresponding
zero contour follows $M_-=0$. The hyperbolic family at small $X$ includes
large negative values of $g_-$. Along the critical diagonal, $g_+$ approaches
$2$ from below, reaching $1.99993$ at $Q/r_0=100$. The detailed limiting
behavior is discussed in Sec.~\ref{sec:extremal}.

\subsection{Mouth asymmetry and the magnetic response}
\label{sec:asym}

Charges, moments and angular momenta defined separately at the two ends of a
wormhole have also been used for rotating charged wormholes constructed
numerically in Einstein--Dirac--Maxwell
theory~\cite{dzhunushalievRotatingWormholesEinsteinDiracMaxwell2026}, and
the gyromagnetic ratios of those numerical solutions have been evaluated at
both ends in Ref.~\cite{Dzhunushaliev:2025ngw}. Here the two ratios follow in
closed form from the static background alone, and the deficit relation fixes
their difference exactly. Cl\'ement's gyromagnetic ratio for a ring
wormhole refers to a geometry with a single flat
end~\cite{clementSelfgravitatingCosmicRings1999}; in the present family
both ends are flat and each carries its own ratio.

Subtracting the two deficit relations gives
\begin{equation}
g_+-g_-=\frac{2(D/r_0)^2\,\mathcal B\,\bigl(w^2-1\bigr)}{w+3q^2\mathcal B} .
\label{eq:gdiff}
\end{equation}
Its sign is the sign of $w-1$, because $\mathcal B$ and $w$ are positive.
On the trigonometric branch, $w=c_-/c_+$ with
$c_\pm=\cos(\alpha\pm\delta)$ and $\alpha=\pi\eta/2$, so
\begin{equation}
w-1=\frac{2\sin\alpha\,\sin\delta}{c_+} .
\label{eq:wminus1}
\end{equation}
It follows that $g_+-g_-$ has the sign of $\delta$, and the ratios are equal
only for the reflection-symmetric member $\delta=0$. This result also holds
at $\eta=\tfrac12$. At that value, $g_+=g_-=1.4591$ for $\delta=0$, whereas
$g_+-g_-=0.664$ for $\delta=0.3$. Since $w^2=G_-^{-1}$,
Eq.~\eqref{eq:gdiff} expresses the difference between the two ratios through
the relative redshift of the two asymptotic clocks: the mouth whose clock runs
slower, measured from the other end, has the smaller gyromagnetic ratio.

\begin{figure*}[!tp]
\centering
\includegraphics[width=\textwidth]{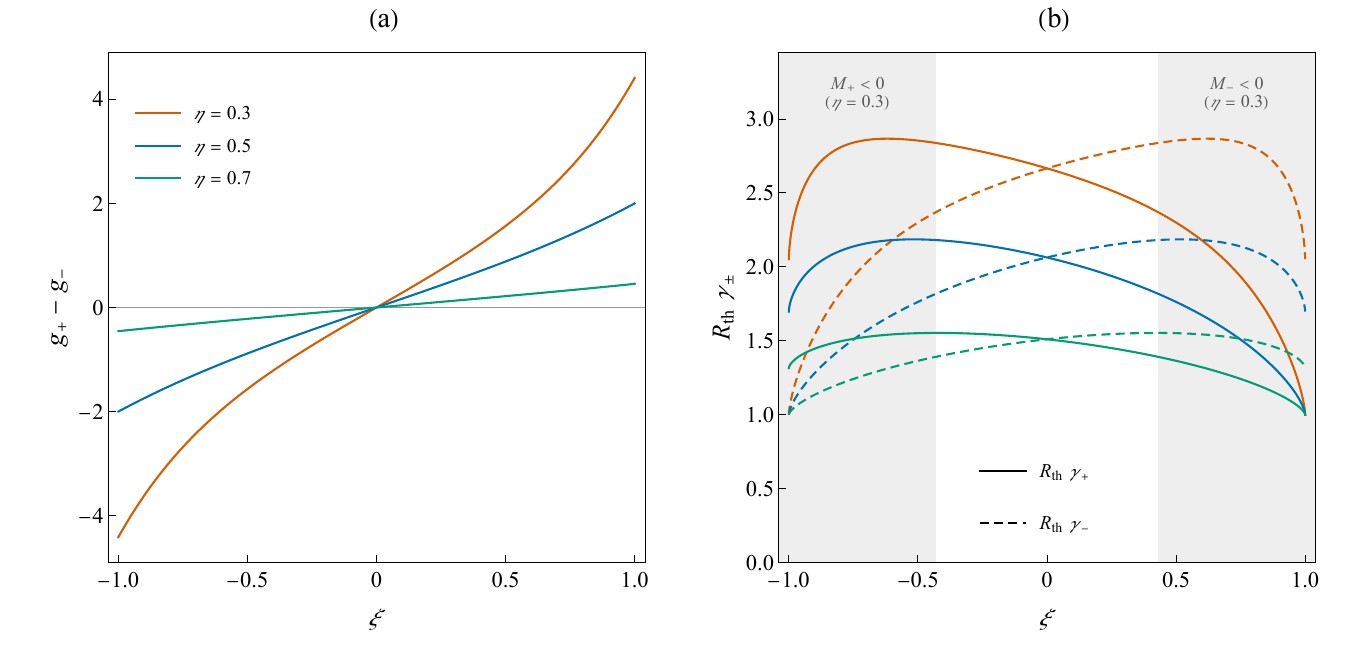}
\caption{Mouth asymmetry on the trigonometric branch for $\eta=0.3$, $0.5$ and
$0.7$, plotted against $\xi=2\delta/[\pi(1-\eta)]\in(-1,1)$. (a) The difference
$g_+-g_-$, which is odd in $\xi$ and vanishes only at reflection symmetry.
(b) The dimensionless response coefficients $R_{\rm th}\gamma_+$ (solid) and
$R_{\rm th}\gamma_-$ (dashed), where $\gamma_s=\mu_s/(Q_sJ_s)$. Exchanging the
mouths corresponds to $\xi\to-\xi$. Shading marks the negative-mass regions for
$\eta=0.3$: $M_+<0$ for $\xi<-3/7$ and $M_-<0$ for $\xi>3/7$. The response
coefficients remain positive across both mass-zero boundaries and within the
shaded regions.}
\label{fig:asym}
\end{figure*}

Figure~\ref{fig:asym} uses $\xi=2\delta/[\pi(1-\eta)]\in(-1,1)$ to compare
three trigonometric sequences. Panel (a) shows the odd dependence of
$g_+-g_-$ on $\xi$, while panel (b) shows $R_{\rm th}\gamma_\pm$.
Since $M_\pm=Q\sin(\alpha\pm\delta)$, negative-mass mouths occur only for
$\eta<\tfrac12$. Of the plotted sequences, only $\eta=0.3$ enters such
regions: $M_+<0$ for $\xi<-3/7$ and $M_-<0$ for $\xi>3/7$.
The response coefficients remain positive across the mass-zero boundaries
and throughout these shaded bands, as Eq.~\eqref{eq:gamma} requires. The limits $\xi\to\pm1$ are the edges
of the regular family, with $Q/R_{\rm th}\to1$ at both.

The sign of the dipole moment must be read together with the charge and
angular-momentum conventions. Since $Q_s=sQ$ and $J_s=-sC/6$, the product
$Q_sJ_s=QJ_+$ is the same at both ends. Thus $\gamma_s>0$ gives the same sign
of $\mu_s$ at the two mouths, although their angular momenta have opposite
signs. The change in sign of $g_s$ across a mass-zero boundary reflects the
factor $M_s$ in Eq.~\eqref{eq:g-gamma}.

\subsection{Field profiles and angular momentum}
\label{sec:fields}

\begin{figure*}[!tp]
\centering
\includegraphics[width=\textwidth]{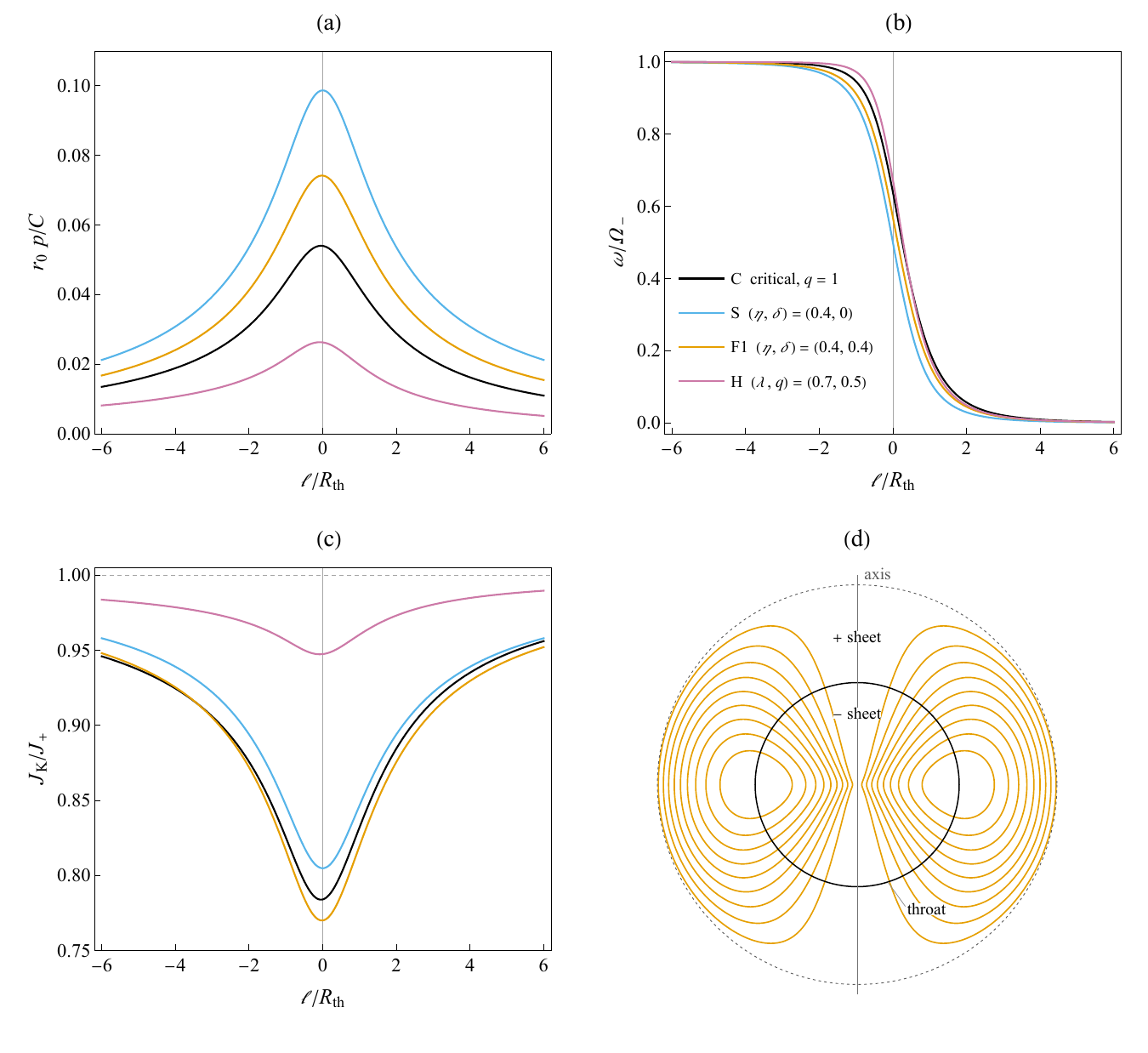}
\caption{Field profiles for \textsf{C} (critical, $q=1$), \textsf{S} (trigonometric,
$(\eta,\delta)=(0.4,0)$), \textsf{F1} (trigonometric, $(\eta,\delta)=(0.4,0.4)$) and \textsf{H}
(hyperbolic, $(\lambda,q)=(0.7,0.5)$), where $q=Q/r_0$. Panels (a)--(c) show
$r_0p/C$, $\omega/\Omega_-$ and $J_{\rm K}/J_+=1-4Qp/C$, respectively, with
$C=-6J_+$. The horizontal coordinate is the signed proper distance from the
throat, $\ell/R_{\rm th}$, positive on the $+$ sheet. The dashed line in (c)
marks unity. Panels (a)--(c) show $|\ell|\le6R_{\rm th}$. (d) Magnetic field lines of \textsf{F1} seen by zero-angular-momentum
observers (ZAMOs), drawn as contours of $\widehat\Psi=(r_0p/C)\sin^2\theta$ at
$\widehat\Psi/\widehat\Psi_{\max}=0.1,\ldots,0.9$. The compactified coordinates
are $\rho=1+(2/\pi)\arctan(\ell/R_{\rm th})$ and
$(x,z)=(\rho\sin\theta,\rho\cos\theta)$, with the left half obtained by
reflection about the rotation axis. This is a coordinate diagram, not an
isometric embedding. The solid circle $\rho=1$ marks the throat, with the
negative sheet inside and the positive sheet outside. In panel (d) the window is
$|\ell|\le15R_{\rm th}$; the dashed circle marks its outer boundary.}
\label{fig:profiles}
\end{figure*}

The frame-dragging profile has a definite direction throughout the wormhole.
To see this, we return to the first-order equations~\eqref{eq:firstorder}
of Sec.~\ref{sec:firstorder}, with $S$ and $H$ as in Eq.~\eqref{eq:SHdefs},
and to the master equation~\eqref{eq:master}, recalling that $C=-6J_+$.
For nonzero spin, define
\begin{equation}
v=1-\frac{4Qp}{C},\qquad \omega'=\frac{Cv}{H^2} .
\label{eq:vdef}
\end{equation}
The master equation then implies
\begin{equation}
\begin{aligned}
(Gv')'-Uv&=-\frac{2G}{S},\\
\bigl[G(1-v)'\bigr]'-U(1-v)&=-\frac{4Q^2G^2}{S^2} .
\end{aligned}
\label{eq:vequations}
\end{equation}
For $Q\ne0$, both sources are strictly negative. A nonpositive interior
minimum of $v$ would have $v'=0$ and $v''\ge0$, making the left-hand side of
the first equation nonnegative. This contradicts its source. Together with
$v\to1$ at both ends, this gives $v>0$. Applying the same argument to $1-v$,
which tends to zero at both ends, gives $v<1$. Hence $0<v<1$ at every finite
$r$. For the neutral solution, $p=0$ and $v=1$.

With $\omega(+\infty)=0$ and $\omega(-\infty)=\Omega_-$, integration yields
\begin{equation}
\frac{\Omega_-}{J_+}=6\int_{-\infty}^{\infty}v\,\frac{G^2}{S^2}\,dr>0 .
\label{eq:Omega-integral}
\end{equation}
The relative rotation therefore has the sign of $J_+$, measured using the
positive-end time. Since $\omega'=Cv/H^2$ and $C=-6J_+$,
$\omega/\Omega_-$ decreases monotonically from $1$ to $0$ as $r$ increases,
for either spin orientation.

The same function $v$ determines the Komar angular
momentum~\cite{Komar:1958wp} on a sphere of constant $r$. Choosing the surface normal toward increasing $r$ gives
\begin{equation}
\begin{aligned}
J_{\rm K}(r)&=-\frac{H^2\omega'}{6}=J_+v,\\
J_{\rm K}-\frac{2}{3}Qp&=J_+ .
\end{aligned}
\label{eq:komar}
\end{equation}
Thus $0<J_{\rm K}/J_+<1$ for $Q\ne0$, irrespective of the sign of $J_+$;
for $Q=0$ the ratio is unity. The normal used here is opposite to the
outward normal at the negative end. The variation of $J_{\rm K}$ along the
wormhole is balanced by the electromagnetic contribution, leaving the
combination in Eq.~\eqref{eq:komar} constant. Because the phantom scalar
does not contribute to $T^t{}_\varphi$, this balance is purely
electromagnetic: the share of the angular momentum not seen by the Komar
integral on a given sphere, $-\tfrac23Qp$, is carried by the Maxwell field, and
it is strictly positive for every charged configuration. For Kerr--Newman it
equals $\tfrac23Q\mu/R$, the flat-space value of the field angular momentum
outside a sphere of radius $R$. Appendix~\ref{app:komar}
derives the surface integral and checks the conventions against the
slowly rotating Kerr--Newman solution.

Figure~\ref{fig:profiles}(a)--(c) compares the four configurations marked in
Fig.~\ref{fig:gmaps}. Their masses, charges and throat radii differ. The magnetic potential
has broader tails than the frame-dragging transition: $p$ and $1-v$ decay as
$1/|r|$, whereas $\omega=O(r^{-3})$ at the positive end and
$\omega-\Omega_-=O(|r|^{-3})$ at the negative end.

Panel (d) shows the \textsf{F1} magnetic field measured by zero-angular-momentum
observers on a constant-$t$ slice. The window $|\ell|/R_{\rm th}\le15$ excludes a small
central disk and a thin outer annulus. Compactification compresses the
negative-end region, producing narrow inner turns in the otherwise smooth
field lines of Fig.~\ref{fig:profiles}(d). The lowest contour crosses the
equator at $\ell/R_{\rm th}=-14.08$ and $+12.46$, showing the asymmetry between
the sheets. The dipole moments also differ: $\mu_+/J_+=0.922$ and
$\mu_-/J_+=1.114$.

\subsection{Relative frame dragging between the ends}
\label{sec:framedrag}

\begin{figure*}[!tp]
\centering
\includegraphics[width=\textwidth]{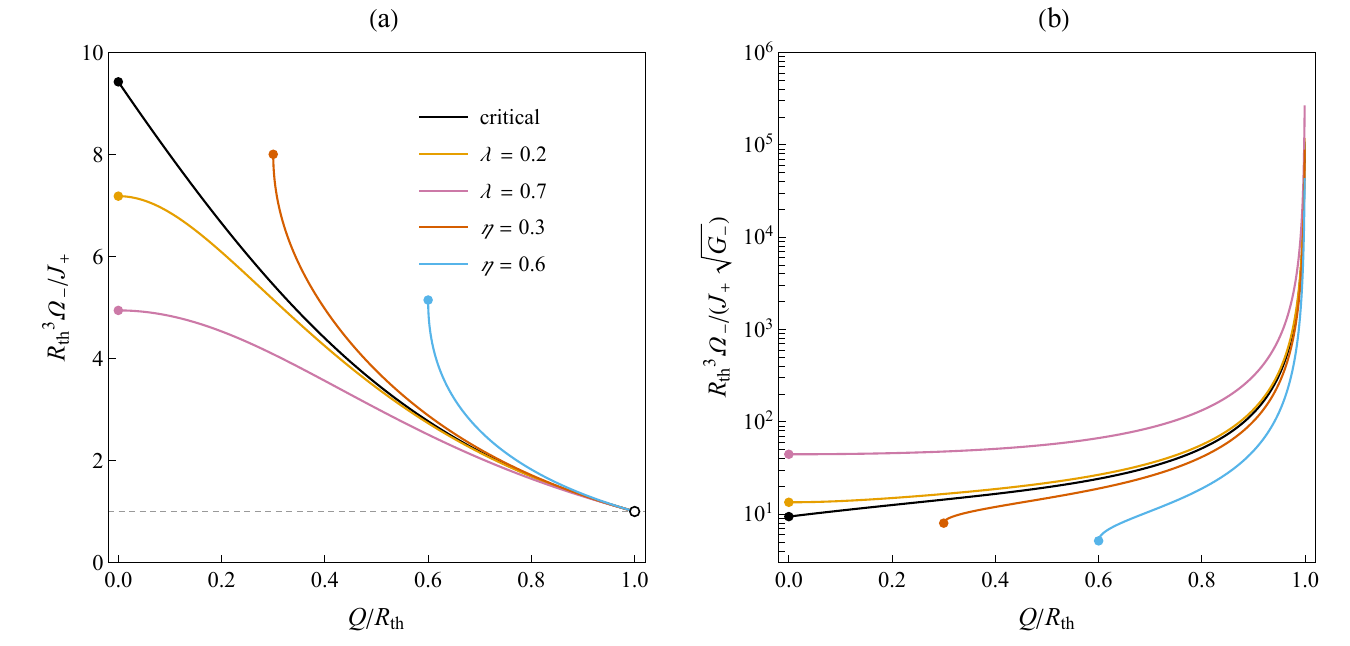}
\caption{Relative rotation of the asymptotic frames along five sequences at
fixed $D/r_0>0$: critical, hyperbolic with $\lambda=0.2$ and $0.7$, and
trigonometric with $\eta=0.3$ and $0.6$. Panel (a) shows
$R_{\rm th}^3\Omega_-/J_+$, using positive-end coordinate time. Panel (b) shows
$R_{\rm th}^3\Omega_-/(J_+\sqrt{G_-})$, using the negative-end proper-time
rate, on a logarithmic scale. Both use $J_+$ as the reference angular
momentum. Filled circles mark neutral endpoints for the critical and
hyperbolic sequences, and charged symmetric starts $\delta=0$,
$Q/R_{\rm th}=\eta$, for the trigonometric sequences. The neutral values in (a)
recover the Kashargin--Sushkov result~\cite{kasharginSlowlyRotatingWormholes2008}, including
$3\pi$ for the critical case. Curves stop at $Q/R_{\rm th}=0.999$. The dashed
line and open circle in (a) mark the limit $1$ along these
paths (Sec.~\ref{sec:extremal}). The growth in (b) reflects
$G_-\to0$.}
\label{fig:framedrag}
\end{figure*}

The integral in Eq.~\eqref{eq:Omega-integral} has the closed form
\begin{equation}
\frac{r_0^3\,\Omega_-}{J_+}
=\frac{3\,s_\pi\,\mathcal B}{w\bigl(w+3q^2\mathcal B\bigr)},
\qquad s_\pi=S_a(\pi) .
\label{eq:Omega-closed}
\end{equation}
Figure~\ref{fig:framedrag} compares the dimensionless quantities
\begin{equation}
\begin{aligned}
A&=\frac{R_{\rm th}^3\,\Omega_-}{J_+}
=\Bigl(\frac{R_{\rm th}}{r_0}\Bigr)^3\frac{r_0^3\,\Omega_-}{J_+},\\
A_{\rm prop}&=\frac{A}{\sqrt{G_-}} .
\end{aligned}
\label{eq:Adef}
\end{equation}
The five sequences keep $D/r_0$ fixed and nonzero. The critical sequence \textsf{C}
and the two hyperbolic sequences, with $\lambda=0.2$ and $0.7$, start at
neutral wormholes. After the throat scaling of Eq.~\eqref{eq:Adef}, their
endpoints recover the first-order result of Kashargin and
Sushkov~\cite{kasharginSlowlyRotatingWormholes2008}, whose $r_0$-normalized form is given in
Table~\ref{tab:limits}. At the neutral endpoint of the critical sequence,
both ordinates equal $3\pi\simeq9.4248$. The trigonometric sequences, with
$\eta=0.3$ and $0.6$,
instead start at charged symmetric configurations. There $X=\eta$ and
$G_-=1$, so the two ordinates coincide, taking the values $8.0039$ and
$5.1466$, respectively. At fixed $\eta>0$ these sequences have no neutral
endpoint.

The vertical tangents at the trigonometric starting points follow from the
different responses of $X$ and $A$ to a small asymmetry. At fixed $\eta$,
\begin{equation}
\begin{aligned}
X&=\eta+\frac{\eta\,\delta^2}{2(1-\eta^2)}+O(\delta^4),\\
\left.\frac{\partial\ln A}{\partial\delta}\right|_{\delta=0}
&=-\tan\alpha,\\
\left.\frac{\partial\ln A_{\rm prop}}{\partial\delta}\right|_{\delta=0}
&=+\tan\alpha .
\end{aligned}
\label{eq:tangents}
\end{equation}
Thus $X$ changes quadratically in $\delta$, while the two ordinates change
linearly and in opposite directions. Exchanging the mouths gives the exact
relation $A(-\delta)=A_{\rm prop}(\delta)$, connecting panel (b) to the
positive-end rate of the mirror configuration.

On the sampled sequences, $A$ decreases and $A_{\rm prop}$ increases with
$X$, while the radial monotonicity of $\omega$ holds analytically on every
background, Eq.~\eqref{eq:vequations}. Along these paths, $A\to1$ as
$X\to1$, with values between $1.00295$ and $1.00299$ at $X=0.999$.
Section~\ref{sec:extremal} compares this behavior with the different limit
along the symmetric trajectory. Meanwhile $G_-\to0$, so
$A_{\rm prop}=A/\sqrt{G_-}$ diverges. This growth records the unbounded
relative redshift of the negative-end clock: the same relative rotation,
measured in the proper time of the negative mouth, becomes arbitrarily fast as
that mouth approaches its extremal limit.

\section{Limiting behavior}
\label{sec:extremal}

Table~\ref{tab:limits} collects the exact limits of the response. Along the
positive-mouth trajectories E1--E4 the positive-end ratio approaches the
Kerr--Newman value cubically, with a coefficient that depends on the
trajectory. Along the neutral trajectories N1--N2 the ratios tend to finite
values that are not 2. Every entry is dimensionless and independent of $r_0$,
and every statement below refers to a named trajectory.

\begin{table*}
\caption{\label{tab:limits}Limiting trajectories. Entries are dimensionless and independent of $r_0$; $\epsilon\equiv1-Q/R_{\rm th}$, $q=Q/r_0$, and $\Omega_-$ is the rotation rate of the negative-end frame in $+$-end coordinate time. Upper block: near-extremal trajectories toward the positive mouth. Along the specified positive-mouth trajectories at fixed $D/r_0>0$ (E1--E3), $(2-g_+)/\epsilon^3\to\tfrac{16}{27}$ and $\Omega_-M_+^3/J_+\to1$; along the symmetric trajectory (E4), on which $D/r_0\to0$, the limits are $\tfrac{\pi^2}{3}$ and 2. On all four, $(2-g_+)Q^2/D^2\to\tfrac23$, the bound of Eq.~\eqref{eq:bound}. Lower block: neutral trajectories ($Q\to0$ at fixed background class).}
\begin{ruledtabular}
\begin{tabular}{llccc}
\multicolumn{5}{l}{\emph{Near-extremal trajectories} ($\epsilon\to0$)}\\
 & Trajectory (held fixed) & $D/r_0$ along it & $\lim(2-g_+)/\epsilon^3$ & $\lim\Omega_-M_+^3/J_+$\footnotemark[1]\\
\colrule
E1 & critical, $q\to\infty$ & $1$ & $16/27$ & $1$\\
E2 & \cellstack[l]{hyperbolic, fixed $\lambda>0$,\\ $q\to\infty$} & $\sqrt{1+\lambda^2}$ & $16/27$ & $1$\\
E3 & \cellstack[l]{trigonometric, fixed $0<\eta<1$,\\ $\delta\to\big[\tfrac{\pi}{2}(1-\eta)\big]^-$\footnotemark[2]} & $\sqrt{1-\eta^2}$ & $16/27$ & $1$\\
E4 & symmetric, $\delta=0$, $\eta\to1$\footnotemark[3] & $\sqrt{1-\eta^2}\to0$ & $\pi^2/3$ & $2$\\
\colrule
\multicolumn{5}{l}{\emph{Neutral trajectories} ($Q\to0$)\footnotemark[4]}\\
 & Trajectory (neutral background) & $\lim g_+$ & $\lim g_-$ & $\lim r_0^3\Omega_-/J_+$\footnotemark[5]\\
\colrule
N1 & \cellstack[l]{critical, $q\to0$\\ (massless Ellis--Bronnikov)} & $0$ & $0$ & $3\pi$\\
N2 & \cellstack[l]{hyperbolic, fixed $\lambda>0$, $q\to0$\\ (massive Ellis--Bronnikov)} & $\dfrac{1+7\lambda^2-(1+\lambda^2)e^{-2\pi\lambda}}{1+4\lambda^2}$ & $2-\dfrac{(1+\lambda^2)\big(1+e^{2\pi\lambda}\big)}{1+4\lambda^2}$ & $\dfrac{3\big(1-e^{-4\pi\lambda}\big)}{4\lambda(1+4\lambda^2)}$\\
\end{tabular}
\end{ruledtabular}
\footnotetext[1]{Since $M_+/R_{\rm th}\to1$ on these trajectories, $M_+^3$ may be replaced by $R_{\rm th}^3$.}
\footnotetext[2]{The positive-mouth edge; $\eta=\tfrac12$ is included (regular representation of the trigonometric branch).}
\footnotetext[3]{Here $\epsilon=1-\eta$ exactly, $2-g_+=\tfrac{\pi^2}{3}\epsilon^3+\tfrac{\pi^2}{2}\epsilon^4+O(\epsilon^5)$, and $\Omega_-M_+^3/J_+=6\eta^2\sin^4\!\alpha\cos\pi\eta/[(1+2\eta^2)\cos^2\!\alpha-3\eta^2]$ with $\alpha=\pi\eta/2$.}
\footnotetext[4]{$g_\pm$ is undefined at $Q=0$; the displayed limits follow from the linear electromagnetic response about the neutral background. N2 reduces to N1 as $\lambda\to0$. At fixed $\eta>0$ the trigonometric branch cannot reach $Q=0$, since $Q=\eta r_0/\cos(\alpha+\delta)\ge\eta r_0$.}
\footnotetext[5]{The limit equals the value computed directly at $Q=0$ (Kashargin--Sushkov).}
\end{table*}

\subsection{Approach to extremality}
\label{sec:nearext}

We measure the distance from extremality by
\begin{equation}
\epsilon\equiv1-\frac{Q}{R_{\rm th}} .
\label{eq:epsdef}
\end{equation}
The limit $\epsilon\to0$ is reached at both edges of the trigonometric family,
so it does not by itself single out a mouth. At the negative-mouth edge the
roles are exchanged: for $\eta=\tfrac12$ and $\delta\to(-\pi/4)^+$,
Eq.~\eqref{eq:deficit} gives $g_+\to0$ and $g_-\to2$. We therefore follow the
positive mouth along the four trajectories E1--E4 of Table~\ref{tab:limits}:
the critical and hyperbolic families as $q\to\infty$, the trigonometric family
at fixed $\eta$ toward its positive-mouth edge, and the symmetric path
$\delta=0$, $\eta\to1$. The statements for $g_-$ at the opposite edge follow by
the mouth exchange $\delta\to-\delta$. Along these trajectories $g_+$
approaches $2$ and the deficit closes as a cubic,
\begin{equation}
\begin{gathered}
2-g_+=\mathcal C_0\,\epsilon^3+O(\epsilon^4),\\
\mathcal C_0=
\begin{cases}
16/27, & \text{E1--E3 (fixed } D/r_0>0),\\
\pi^2/3, & \text{E4 (}D/r_0\to0).
\end{cases}
\end{gathered}
\label{eq:cubic}
\end{equation}

On E1--E3 the quantity $a=k/r_0^2$ is fixed,
$1+a=(D/r_0)^2>0$, and $m\equiv M_+/r_0=\sqrt{q^2+a}\sim q$. Since
$w=C_a(\pi)+mS_a(\pi)$ grows only linearly in $q$, the deficit relation reduces
to $2-g_+\simeq2(1+a)/(3q^2)$, independent of $\mathcal B$. The throat moves
toward the positive end: expanding the throat condition
$W_u\sin u-W\cos u=0$ about $u=0$ gives $m(1+a)u_{\rm th}^3\to3$. The throat
identity of Sec.~\ref{sec:background},
$1-Q^2/R_{\rm th}^2=(1+a)\sin^2u_{\rm th}$, then fixes the distance from
extremality. Together,
\begin{equation}
\epsilon\simeq\frac{1+a}{2}\,u_{\rm th}^2,\qquad
\epsilon^3\simeq\frac{9(1+a)}{8q^2},\qquad
2-g_+\simeq\frac{2(1+a)}{3q^2},
\label{eq:leading}
\end{equation}
and the factor $1+a$ cancels from the ratio, leaving $(2/3)/(9/8)=16/27$ on
each of these trajectories, including the removable case $\eta=\tfrac12$. The
symmetric path behaves differently because its throat does not move: it stays
at $u_{\rm th}=\pi/2$, so the small-angle step above does not apply. There
$1+a=2\epsilon-\epsilon^2$ and $q\simeq2/(\pi\epsilon)$, and the same leading
form of the deficit relation gives $2-g_+\simeq(\pi^2/3)\epsilon^3$.

Both coefficients have a common origin. With $X=Q/R_{\rm th}$, the throat
identity~\eqref{eq:throatid} and $R_{\rm th}=r_0W_{\rm th}/\sin u_{\rm th}$,
where $W_{\rm th}=W(u_{\rm th})$, give $D^2/Q^2=(1-X^2)\,G(r_{\rm th})/X^2$, so
Eq.~\eqref{eq:bound} becomes, exactly,
\begin{equation}
2-g_+=\frac{2}{3}\,\frac{(1-X^2)\,G(r_{\rm th})}{X^2}
\left[1+\frac{r_0^2w}{3Q^2\mathcal B}\right]^{-1}.
\label{eq:throatform}
\end{equation}
On all four trajectories $r_0^2w/(Q^2\mathcal B)\to0$, and therefore
\begin{equation}
2-g_+\simeq\frac{4}{3}\,\epsilon\,G(r_{\rm th}) .
\label{eq:universal}
\end{equation}
The cubic law is thus the statement that the redshift factor at the throat,
measured in positive-end time, vanishes as $\epsilon^2$: on E1--E3
$W_{\rm th}^2\simeq9/(4\epsilon^2)$, and on E4 $W_{\rm th}^2\simeq4/(\pi^2\epsilon^2)$,
which give $16/27$ and $\pi^2/3$. The coefficient depends on the path because
the prefactor of $\epsilon^2$ in $G(r_{\rm th})$ does. Along the path
$\eta=1-s$, $\delta=\pi s/4$, $s\to0$, for example, the same form gives
$\mathcal C_0=\pi^2/(12\sin^6u_*)\simeq1.4036$, where
$u_*-\sin u_*\cos u_*=\pi/4$. We have not shown that $G(r_{\rm th})=O(\epsilon^2)$
on every approach to the edge.

The next order shows that only the leading coefficient is shared. On a
fixed-$a$ trajectory,
\begin{equation}
\begin{gathered}
\frac{2-g_+}{\epsilon^3}=\frac{16}{27}\bigl[1+K(a)\,\epsilon+O(\epsilon^{3/2})\bigr],\\
K(a)=\frac{3(7+3a)}{10(1+a)} ,
\end{gathered}
\label{eq:firstcorr}
\end{equation}
while on E4 the compensated deficit is
$(2-g_+)/\epsilon^3=\pi^2/3+(\pi^2/2)\epsilon+O(\epsilon^2)$. Both corrections are
positive, so every curve approaches its limit from above sufficiently close to
the edge. Equation~\eqref{eq:firstcorr} also explains why the symmetric value
is not a special case of the fixed-$a$ result: $K(a)$ contains $1/(1+a)$, the
expansion is not uniform as $a\to-1$, and the symmetric limit cannot be
obtained by letting $a\to-1$ after taking the fixed-$a$ near-extremal limit.
The full expansions are given in Appendix~\ref{app:limits}.

\begin{figure*}[!tp]
\centering
\includegraphics[width=\textwidth]{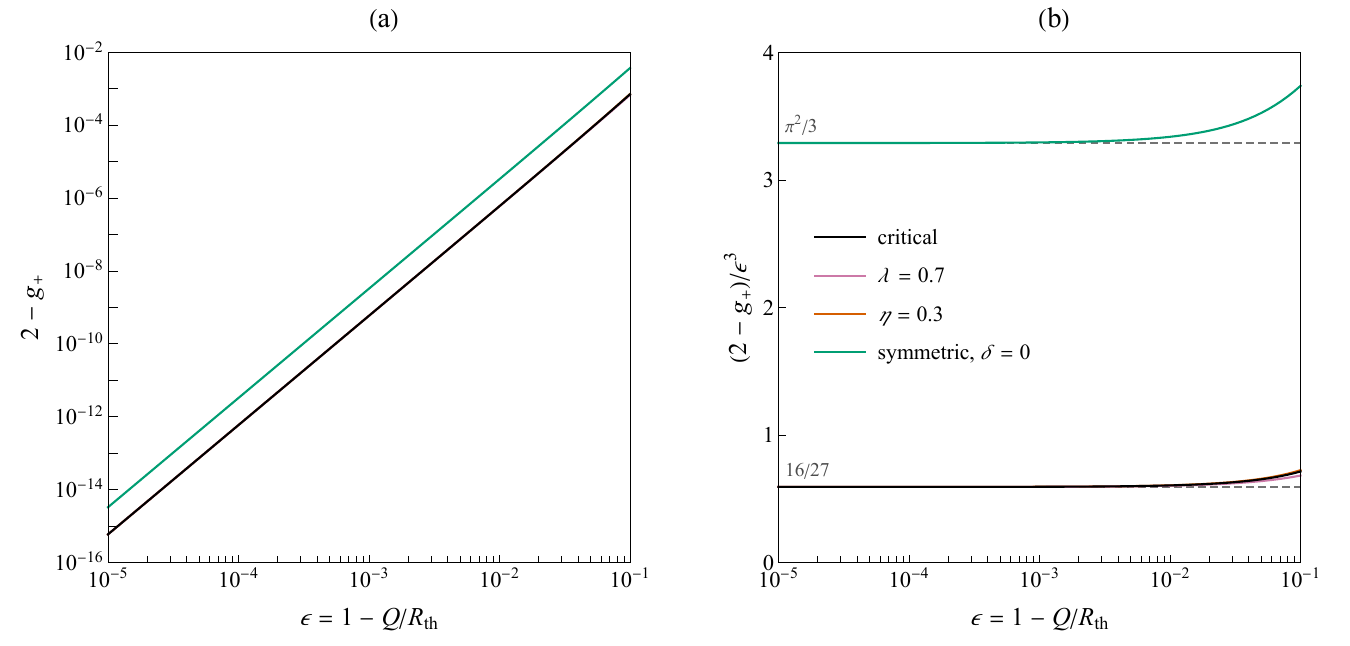}
\caption{Path-dependent near-extremal scaling of the positive-end deficit,
with $\epsilon\equiv1-Q/R_{\rm th}$; both horizontal axes are logarithmic.
(a) $2-g_+$ on a logarithmic scale. (b) $(2-g_+)/\epsilon^3$ on a linear
scale. Three paths hold $D/r_0>0$ fixed: critical ($q\to\infty$), hyperbolic
with $\lambda=0.7$ ($q\to\infty$) and trigonometric with $\eta=0.3$ (toward the
positive-mouth edge). The fourth is the symmetric path $\delta=0$, $\eta\to1$,
along which $D/r_0\to0$. The dashed lines mark $16/27$, the limit along the
three fixed-$D/r_0$ paths shown, and $\pi^2/3$, the limit along the symmetric
path. The three fixed-$D/r_0$ curves overlap within the line width for
$\epsilon\lesssim10^{-2}$.}
\label{fig:nearext}
\end{figure*}

Figure~\ref{fig:nearext} shows the approach on four representative paths
over $10^{-5}\le\epsilon\le10^{-1}$. Panel (a) displays the cubic over about
twelve decades of the deficit; panel (b) shows the compensated deficit
settling onto $16/27$ for the three fixed-$D/r_0$ paths and onto $\pi^2/3$
for the symmetric path. On the sampled paths the approach is monotonic over
the whole range, and the relative departure from the limit stays below
$2.3\times10^{-3}$ for $10^{-5}\le\epsilon\le10^{-3}$, in agreement with
the exact limits of Table~\ref{tab:limits}.

The frame dragging shows the same path dependence. Along E1--E3,
$\Omega_-M_+^3/J_+\to1$, while along E4 the limit is $2$. Since
$M_+/R_{\rm th}\to1$ on these trajectories, this is also the limit of the
ordinate of Fig.~\ref{fig:framedrag}(a): the E1--E3 limit $1$ is the open
circle at unity there, since all five sequences of that figure lie on fixed
$D/r_0>0$ paths. The E4 limit $2$ is a separate result for the symmetric
path.

The static positive exterior also has a definite limit after rescaling by the
mass. With $\chi=r/M_+$ and $\tau=t/M_+$ at fixed $\chi>0$, one finds
$W\to1+1/\chi$ and $S/M_+^2\to\chi^2$ along E1--E4, so that
\begin{equation}
\begin{aligned}
\frac{ds^2}{M_+^2}\longrightarrow{}
&-\Bigl(1+\frac1\chi\Bigr)^{-2}d\tau^2\\
&+\Bigl(1+\frac1\chi\Bigr)^{2}\bigl(d\chi^2+\chi^2d\Omega^2\bigr),
\end{aligned}
\label{eq:RNlimit}
\end{equation}
the extremal Reissner--Nordstr\"om exterior in isotropic coordinates
(Appendix~\ref{app:limits}). At finite $\epsilon$ every admissible
background remains regular, horizonless and two-ended, so the positive mouth
approaches the extremal Reissner--Nordstr\"om exterior without ever forming a
horizon. This gives a geometric reading of the approach $g_+\to2$: seen from
the positive end, the wormhole becomes indistinguishable at scales of order
$M_+$ from an extremal charged black hole, whose slowly rotating extension is
the Kerr--Newman metric, while the throat recedes into a region where the
redshift factor vanishes, Eq.~\eqref{eq:universal}. The pattern parallels
rapidly rotating neutral Ellis--Bronnikov wormholes, which approach the
extremal Kerr solution as the throat velocity approaches that of
light~\cite{kleihausRotatingEllisWormholes2014,chewGeometrySpinningEllis2016}.

\subsection{Neutral limits}
\label{sec:neutral}

Switching off the charge at fixed $r_0$, $q=Q/r_0\to0$, leads to a different
picture. The frame dragging changes continuously: its limit equals the value
computed directly on the neutral background,
\begin{equation}
\left.\frac{r_0^3\,\Omega_-}{J_+}\right|_{Q=0}
=6\int_0^\pi W_0^{-4}\sin^2u\,du ,
\label{eq:neutralquad}
\end{equation}
which reproduces the first-order results of Kashargin and
Sushkov~\cite{kasharginSlowlyRotatingWormholes2008}: $3\pi$ for the massless Ellis--Bronnikov
wormhole (N1) and the closed form of N2 for the massive one. In
Fig.~\ref{fig:framedrag} these values appear, after the throat scaling, as the
neutral endpoints of the critical and hyperbolic sequences.

The gyromagnetic ratios behave differently. Since $g_s=2\mu_sM_s/(Q_sJ_s)$, they
are undefined at $Q=0$. Their limits are nevertheless finite: at fixed spin,
$\mu_s/Q_s$ tends to the linear response to charge about the neutral
background, and the resulting values are the entries of the lower block
of Table~\ref{tab:limits}. For the massless wormhole both ratios tend to zero. For
the massive one $g_+$ tends to a value strictly below 2, while the far mouth,
which has negative mass, has a negative limit $g_-$, as the sign rule of
Sec.~\ref{sec:deficit} requires. This is the opposite regime to
Eq.~\eqref{eq:universal}: the redshift factor at the throat stays finite as
the charge is switched off, and the ratios tend to the values in
Table~\ref{tab:limits}, which differ from $2$.

Not every family reaches the neutral background. At fixed $\eta>0$ the
trigonometric configurations satisfy $Q=\eta r_0/\cos(\alpha+\delta)\ge\eta
r_0$, which is why the trigonometric sequences of Fig.~\ref{fig:framedrag} have
no neutral endpoint. Trigonometric paths that take $\eta\to0$ at fixed
admissible $\delta$ do reach it; the symmetric path $\delta=0$, for instance,
arrives at the same neutral endpoint as N1, with $g\sim\pi^2\eta^2\to0$.

\subsection{Comparison with black holes}
\label{sec:compare}

The two limits place the wormhole response beside two familiar black-hole
results. The value $g=2$ of the Kerr--Newman
family~\cite{newmanMetricRotatingCharged1965,carterGlobalStructureKerr1968} is
not special to that solution: it holds for slightly charged, stationary, \rev{axisymmetric} black
holes~\cite{waldBlackHoleUniform1974} and for the electrovac fields generated
by Ernst's prescription~\cite{reinaGyromagneticRatioEinsteinMaxwell1975}. Along the positive-mouth trajectories of
Sec.~\ref{sec:extremal}, $g_+$ approaches this value, because the redshift
factor at the throat vanishes there, Eq.~\eqref{eq:universal}.

The slowly rotating dilaton black holes of Horne and
Horowitz~\cite[Eq.~(40)]{horneRotatingDilatonBlack1992} offer an instructive contrast. \rev{To first order in the rotation and at fixed
nonzero dilaton coupling $\alpha_{\rm d}$,} they obey
$2-2\alpha_{\rm d}^2/(\alpha_{\rm d}^2+3)\le g\le2$; \rev{the first-order deficit} stays finite
at extremality, and $g\to2$ as the charge is switched off. (At
$\alpha_{\rm d}=0$ the family reduces to
Kerr--Newman, with $g=2$ throughout.) The wormhole family shows the reverse
pattern: the deficit closes at the positive-mouth edge, cubically, and on the
massive neutral limit N2 the ratios are not $2$ (on N1 both masses vanish, and
so does $g_s$). The contrast has a simple origin. The dilaton charge of those
black holes is secondary hair, proportional to $Q^2/M$, so it switches off with
the charge and $g\to2$ in the neutral limit is automatic. Here the scalar
charge $D$ is primary hair, independent of $Q$, so the deficit cannot close as
$Q\to0$. Nor does a scalar lower the gyromagnetic ratio in general: in
Einstein--Maxwell-scalar theories $g$ depends on the
couplings~\cite{qiuSlowlyRotatingBlack2021}.

The first-integral construction used here parallels that of
Qiu~\cite{qiuSlowlyRotatingBlack2021} for slowly rotating black holes in an
Einstein--Maxwell-scalar theory, in which the first-integral constant is
likewise identified with the angular momentum; here it is adapted to a
geometry with two asymptotic ends, where the same constant fixes the angular
momenta of both.

\section{Conclusions}
\label{sec:conclusions}

We have obtained the slowly rotating charged Ellis--Bronnikov wormholes in
closed form on all three background branches and shown that, within the
stationary axial dipole sector and with no applied magnetic field, the
first-order solution is unique. Because the geometry has two asymptotically
flat ends, its gyromagnetic ratio is a pair of numbers, and the exact deficit
relation fixes both of them in terms of the static background alone: the
scalar and electric charges, a single branch factor and the normalization of
each end.

The deficit relation bounds the departure from the Kerr--Newman value by
two-thirds of the squared ratio of scalar to electric charge. Both ratios lie
below 2---automatically at a negative-mass mouth, since each carries the sign
of the mass of its own mouth---and the two coincide only for
reflection-symmetric throats. Their difference is fixed by the relative
redshift of the two asymptotic clocks and is therefore a direct measure of the
asymmetry of the wormhole. The Kerr--Newman value is approached along every
near-extremal trajectory studied, where the redshift factor at the throat
vanishes and the exterior of the positive mouth approaches that of an extremal
Reissner--Nordstr\"om black hole without forming a horizon. The cubic closing
of the deficit and the dependence of its coefficient on the path follow from
how the throat redshift scales with the distance from extremality.

Rotation also gives the wormhole a definite internal kinematics. The two ends
rotate relative to each other in the sense of the angular momentum measured at
the positive mouth, the dragging profile decreases monotonically across the
throat, and on every sphere the Maxwell field carries a strictly positive
share of the angular momentum. When the charge is switched off, the frame
dragging reproduces the known neutral results, whereas on the massive neutral
branch the gyromagnetic ratios remain away from the Kerr--Newman value. This
reverses the behavior of slowly rotating dilaton black holes, whose deficit
survives at extremality and closes when the charge is removed, a difference
that follows from the scalar charge being primary rather than secondary hair.

The first-order solution is the starting point for several extensions. At
second order the rotation deforms the throat and induces quadrupole moments,
which can be computed along the lines of the neutral second-order analyses
discussed in Sec.~\ref{sec:sloweqs}. Rapidly rotating charged solutions would
carry the deficit relation beyond perturbation theory. Finally, the static
charged backgrounds are unstable, and rotation is known to slow the radial
instability of neutral Ellis--Bronnikov wormholes. The stationary response
computed here therefore describes configurations whose lifetime is not yet
known; a time-dependent analysis of the slowly rotating charged family is the
natural next step to establish how rotation and charge together modify the
unstable modes.

\appendix
\section{Closed forms on the three branches}
\label{app:branches}

This appendix collects the branch-by-branch expressions behind the unified
results of Secs.~\ref{sec:firstorder} and~\ref{sec:results}: the background in
a common form, the throat chart used in Fig.~\ref{fig:gmaps}, the rotation
solutions, and the moments, gyromagnetic ratios and frame-dragging rates on
each branch.

\subsection{Background}

We use the unified form $W=C_a(u)+m\,S_a(u)$ of Eq.~\eqref{eq:Wunified},
with $C_a$ and $S_a$ from Eq.~\eqref{eq:CS} and $m=M_+/r_0$, $m^2=q^2+a$.
The sign of $m$ is that of $M_+$. For the
critical and hyperbolic parametrizations below, we choose $Q>0$ and $M_+>0$;
the opposite orientations follow by exchanging the ends and restoring
$G_+=1$. On the trigonometric branch $M_+$ may be negative. The three
parametrizations used in the text are as follows.

\textit{Trigonometric} ($0<\eta<1$, $|\delta|<\tfrac{\pi}{2}(1-\eta)$,
$\alpha=\pi\eta/2$, $c_\pm=\cos(\alpha\pm\delta)$):
\begin{equation}
\begin{gathered}
W=\frac{Q}{\eta r_0}\cos(\alpha+\delta-\eta u),\qquad Q=\frac{\eta r_0}{c_+},\\
M_\pm=Q\sin(\alpha\pm\delta),\qquad w=\frac{c_-}{c_+}.
\end{gathered}
\label{eq:trigbg}
\end{equation}

\textit{Critical}:
\begin{equation}
W=1+qu,\qquad M_\pm=\pm Q,\qquad w=1+\pi q .
\label{eq:critbg}
\end{equation}

\textit{Hyperbolic} ($\lambda>0$, $\sinh\zeta=\lambda/q$,
$\zeta_-=\zeta+\pi\lambda$):
\begin{equation}
\begin{gathered}
W=\frac{q}{\lambda}\sinh(\lambda u+\zeta),\qquad w=\frac{\sinh\zeta_-}{\sinh\zeta},\\
M_+=r_0\sqrt{q^2+\lambda^2},\qquad M_-=-r_0q\cosh\zeta_- .
\end{gathered}
\label{eq:hypbg}
\end{equation}

In each case $w=W(\pi)=G_-^{-1/2}$ and $s_\pi=S_a(\pi)$. The branch factor
$\mathcal B=C_a(\pi)/(1+4a)$ of Eq.~\eqref{eq:Bdef} has a removable singularity
at $\eta=\tfrac12$ on the trigonometric branch. A form regular there is
\begin{equation}
\mathcal B=\frac{\pi}{2(1+2\eta)}\,
\frac{\sin\bigl[\pi(\tfrac12-\eta)\bigr]}{\pi(\tfrac12-\eta)},
\qquad \mathcal B\bigl(\tfrac12\bigr)=\frac{\pi}{4},
\label{eq:Bregular}
\end{equation}
which is used in all numerical evaluations near $\eta=\tfrac12$.

\subsection{Throat chart}

The throat satisfies the condition~\eqref{eq:throatcond} and the
identity~\eqref{eq:throatid} of Sec.~\ref{sec:throat}.
Equation~\eqref{eq:throatid} allows the pair $(X,u_{\rm th})$, with
$X=Q/R_{\rm th}$, to replace the branch parameters. Solving it for $a$, and
Eq.~\eqref{eq:throatcond} for $m$, gives
\begin{equation}
\begin{gathered}
a=\frac{\cos^2u-X^2}{\sin^2u},\qquad
\frac{r_0}{R_{\rm th}}=C_a(u)\sin u-S_a(u)\cos u,\\
Y\equiv\frac{M_+}{R_{\rm th}}=C_a(u)\cos u-a\,S_a(u)\sin u,
\end{gathered}
\label{eq:chart}
\end{equation}
with $u=u_{\rm th}$. Along a line of fixed $X$,
\begin{equation}
\left.\frac{\partial Y}{\partial u}\right|_X
=(1+a)\bigl(C_a\sin u-S_a\cos u\bigr)(u\cot u-1)<0 ,
\label{eq:chartmono}
\end{equation}
because $1+a=(D/r_0)^2>0$, the middle factor is $r_0/R_{\rm th}>0$, and
$u\cot u<1$ on $(0,\pi)$. The map $(X,u_{\rm th})\mapsto(X,Y)$ is therefore
one-to-one: each configuration, up to the overall scale $r_0$, appears once in
the plane of Fig.~\ref{fig:gmaps}. At fixed $X$, decreasing $u_{\rm th}$
runs through the mirror hyperbolic sector ($u_{\rm th}>\pi-\arccos X$), the
opposite critical line $Y=-X$, the trigonometric sector, the critical line
$Y=X$ at $u_{\rm th}=\arccos X$, and the hyperbolic sector
($u_{\rm th}<\arccos X$). As $u_{\rm th}\to0$, $a\to\infty$ with
$a\,u_{\rm th}^2\to1-X^2$, and Eq.~\eqref{eq:chart} gives the upper boundary
$Y_{\max}=\cosh t-t\sinh t$, $t=\sqrt{1-X^2}$, of Eq.~\eqref{eq:Ymax}.

\subsection{Rotation solutions}
\label{app:rotsol}

The rotation functions are written with $y=\arctan(r/r_0)=\pi/2-u$, so that
$\tan y=r/r_0$ and $\sec^2y=S/r_0^2$. On each branch we give two independent
solutions of the homogeneous master equation~\eqref{eq:master} ($C=0$), the
solution $p$ that satisfies Eq.~\eqref{eq:bc}, and $\omega$ with
$\omega(+\infty)=0$. A prime on a homogeneous solution denotes $d/dy$.

\textit{Critical.} One homogeneous solution is
\begin{equation}
p_h=\frac{q^2}{2(1+qu)}-\frac{q}{4}\cot u-\frac{1+qu}{4}\csc^2u ,
\end{equation}
and a second is $p_h\int dr/(Gp_h^2)$, which is elementary. With
$\mathcal D_2=1+\pi q+3q^2$,
\begin{equation}
\begin{gathered}
p=\frac{Cq\csc^2u\,P_{\rm c}(u)}{8r_0\mathcal D_2(1+qu)},\qquad
\omega=\frac{C}{4qr_0^3\mathcal D_2}\Bigl[\frac{\Omega_{\rm c}(u)}{(1+qu)^4}-2\Bigr],\\
\begin{aligned}
P_{\rm c}&=-1-2\pi q+(2\pi+3q)u+2(\pi q-1)u^2-2qu^3\\
&\quad+(1+2\pi q-3qu)\cos2u+(2u-\pi+\pi qu)\sin2u,\\
\Omega_{\rm c}&=2+2q^2+\pi q^3+(6q-2\pi q^2)u+(8q^2-2\pi q^3)u^2\\
&\quad+4q^3u^3-q^2(2+\pi q)\cos2u+q(1+\pi q)\sin2u .
\end{aligned}
\end{gathered}
\label{eq:critsol}
\end{equation}

\textit{Trigonometric.} With $T=\eta y+\delta$, $h=\cos\pi\eta$,
$\hat c=\cos2\delta$ and $\hat s=\sin2\delta$, a homogeneous pair and a
particular solution are
\begin{equation}
\begin{aligned}
h_{\rm o}&=\sec^2y\sin T\cos T+\eta\tan y-2\eta^2\tan T,\\
h_{\rm e}&=\sec^2y\bigl[(1+2\eta^2)\cos^2T-3\eta^2\bigr]\\
&\quad-3\eta^2-6\eta^3\tan y\tan T,\\
h_{\rm o}h_{\rm e}'-h_{\rm o}'h_{\rm e}&=-2\eta(1-\eta^2)(1-4\eta^2)\cos^2T\sec^2y,\\
p_{\rm P}&=\frac{Cc_+}{12\eta^3r_0}\bigl(3\eta^2-\sec^2y\cos^2T\bigr).
\end{aligned}
\label{eq:trigbasis}
\end{equation}
With $\mathcal F=\cos2\delta+(1+2\eta^2)\mathcal B$ and
$\mathcal D=\tfrac12\bigl[(1-4\eta^2)\hat c+(1+2\eta^2)h\bigr]
=\tfrac12(1-4\eta^2)\mathcal F$,
\begin{equation}
\begin{aligned}
p={}&\frac{Cc_+}{4\eta r_0\mathcal D}\Bigl\{\sec^2y\bigl(\cos^2\eta y-\cos^2\alpha\bigr)
+\eta^2(h-2\hat c)\\
&+\eta\hat s\tan y-\eta\bigl[2\eta\hat s+(\hat c+h)\tan y\bigr]\tan T\Bigr\},\\
\omega={}&\frac{Cc_+^4}{2r_0^3\mathcal D}\bigl[\mathcal R(y)-\mathcal R(\tfrac\pi2)\bigr],\\
\mathcal R={}&\tfrac12(h+\hat c)\sin y\cos y\sec^4T+\eta\hat s\cos^2y\sec^4T\\
&+\frac{1}{2\eta}\bigl[(3h-\hat c)\tan T+(h+\hat c)\tan^3T\\
&\qquad-2\hat s\tan^2T\bigr].
\end{aligned}
\label{eq:trigsol}
\end{equation}

At $\eta=\tfrac12$ the chosen basis degenerates, $h_{\rm e}=-\tfrac32\tan2\delta\,h_{\rm o}$,
and $\mathcal D=0$. A second homogeneous solution is
$\tilde h=\lim_{\eta\to1/2}(h_{\rm e}+\tfrac32\tan2\delta\,h_{\rm o})/(1-4\eta^2)$,
with $h_{\rm o}\tilde h'-h_{\rm o}'\tilde h=-\tfrac34\cos^2(\delta+y/2)\sec^2y$; it
contains a term linear in $y$. With $z_0=y/2+\delta$, $T_0=\tan z_0$,
$t=\tan y$ and $\mathcal F_0=\hat c+3\pi/8$, which is $\mathcal F$ at $\eta=\tfrac12$, the
regular expressions are
\begin{equation}
\begin{aligned}
p={}&-\frac{Cc_+}{8\mathcal F_0r_0}\Psi,\qquad
\omega=-\frac{Cc_+^4}{4\mathcal F_0r_0^3}\bigl[\mathcal R_1(y)-\mathcal R_1(\tfrac\pi2)\bigr],\\
\Psi={}&\sec^2y\,(\pi-2y\sin y)-4\hat c-\tfrac{\pi}{2}-4\hat sT_0-\hat s\,y\sec^2z_0\\
&+t\bigl[2\hat s+(\pi-2\hat c)T_0-\hat c\,y\sec^2z_0\bigr],\\
\mathcal R_1={}&\sec^4z_0\bigl[\bigl(-\tfrac{\pi}{2}+2\hat c\,yT_0\bigr)\sin y\cos y\\
&\quad+\hat s(1+2yT_0)\cos^2y\bigr]-\pi(3T_0+T_0^3)\\
&+y\sec^2z_0\bigl[\hat c(3T_0^2-1)-4\hat sT_0\bigr]\\
&+2\hat cT_0-2\hat cT_0^3+4\hat sT_0^2 .
\end{aligned}
\label{eq:halfsol}
\end{equation}

\textit{Hyperbolic.} The continuation $\eta=i\lambda$, $\delta=\pi/2-i\beta$,
with $\beta=\zeta+\pi\lambda/2$, turns the trigonometric expressions into real
ones. With $\psi=\lambda u+\zeta=\beta-\lambda y$,
\begin{equation}
\begin{aligned}
H_{\rm o}&=\tfrac12\sec^2y\sinh2\psi+\lambda\tan y-2\lambda^2\coth\psi,\\
H_{\rm e}&=-\sec^2y\bigl[(1-2\lambda^2)\sinh^2\psi-3\lambda^2\bigr]\\
&\quad+3\lambda^2+6\lambda^3\tan y\coth\psi,\\
H_{\rm o}H_{\rm e}'-H_{\rm o}'H_{\rm e}&=2\lambda(1+\lambda^2)(1+4\lambda^2)\sinh^2\psi\sec^2y,
\end{aligned}
\label{eq:hypbasis}
\end{equation}
so this basis never degenerates. With
\begin{equation*}
\begin{aligned}
\mathcal K&=\tfrac12\bigl[(1+4\lambda^2)\cosh2\beta-(1-2\lambda^2)\cosh\pi\lambda\bigr]\\
&=(1+4\lambda^2)\sinh\zeta\sinh\zeta_-+3\lambda^2\cosh\pi\lambda>0,
\end{aligned}
\end{equation*}
\begin{equation}
\begin{aligned}
p={}&-\frac{C\sinh\zeta}{4\lambda r_0\mathcal K}\Bigl\{\sec^2y\bigl(\cosh^2\lambda y-\cosh^2\tfrac{\pi\lambda}{2}\bigr)\\
&-\lambda^2(\cosh\pi\lambda+2\cosh2\beta)-\lambda\sinh2\beta\tan y\\
&+\bigl[2\lambda^2\sinh2\beta\\
&\quad+\lambda(\cosh2\beta-\cosh\pi\lambda)\tan y\bigr]\coth\psi\Bigr\},\\
\omega={}&-\frac{C\sinh^4\zeta}{2r_0^3\mathcal K}\bigl[\mathcal R_H(y)-\mathcal R_H(\tfrac\pi2)\bigr],\\
\mathcal R_H={}&\frac{(\cosh\pi\lambda-\cosh2\beta)\sin y\cos y}{2\sinh^4\psi}
-\frac{\lambda\sinh2\beta\cos^2y}{\sinh^4\psi}\\
&+\frac{1}{2\lambda}\bigl[(\cosh\pi\lambda-\cosh2\beta)\coth^3\psi\\
&\quad+2\sinh2\beta\coth^2\psi\\
&\quad-(3\cosh\pi\lambda+\cosh2\beta)\coth\psi\bigr].
\end{aligned}
\label{eq:hypsol}
\end{equation}

\subsection{Moments, gyromagnetic ratios and frame dragging}

The dipole moments follow from the magnetic tails through Eq.~\eqref{eq:asym}.
Since $Q_sJ_s=QJ_+$ at both ends, Eq.~\eqref{eq:g-gamma} gives
$g_\pm=2(\mu_\pm/J_+)(M_\pm/Q)$. In unified form, with
$m_-=M_-/r_0=-W_u(\pi)$,
\begin{equation}
\frac{\mu_+}{J_+}=\frac{q\,(s_\pi+3m\mathcal B)}{w+3q^2\mathcal B},\qquad
\frac{\mu_-}{J_+}=\frac{q\,(w\,s_\pi+3m_-\mathcal B)}{w+3q^2\mathcal B}.
\label{eq:muunified}
\end{equation}
These follow from Eq.~\eqref{eq:deficit} and the identities
$w-m\,s_\pi=C_a(\pi)=(1+4a)\mathcal B$ and $w\,C_a(\pi)+m_-s_\pi=1$; they
remain finite where $M_+$ or $M_-$ vanishes. On each branch:

\textit{Critical} ($\mathcal D_2=1+\pi q+3q^2$):
\begin{equation}
\begin{gathered}
\frac{\mu_+}{J_+}=\frac{q(\pi+3q)}{\mathcal D_2},\qquad
\frac{\mu_-}{J_+}=\frac{q\bigl[\pi+(\pi^2-3)q\bigr]}{\mathcal D_2},\\
2-g_+=\frac{2}{\mathcal D_2},\qquad 2-g_-=\frac{2(1+\pi q)^2}{\mathcal D_2},\\
\frac{r_0^3\Omega_-}{J_+}=\frac{3\pi}{(1+\pi q)\mathcal D_2}.
\end{gathered}
\label{eq:critforms}
\end{equation}

\textit{Trigonometric} ($\mathcal F$ as above,
$z_\pm=\delta\pm\alpha$):
\begin{equation}
\begin{gathered}
\frac{\mu_\pm}{J_+}=\frac{2}{\mathcal F}\bigl[c_\pm\sin\pi\eta\pm3\eta^2\mathcal B\sin z_\pm\bigr],\\
2-g_\pm=\frac{4(1-\eta^2)\mathcal B\,c_\pm^2}{\mathcal F},\qquad
\frac{r_0^3\Omega_-}{J_+}=\frac{6c_+^3\mathcal B\sin\pi\eta}{\eta\,c_-\mathcal F}.
\end{gathered}
\label{eq:trigforms}
\end{equation}

\textit{Hyperbolic} ($\mathcal K$ as above, $\zeta_+\equiv\zeta$):
\begin{equation}
\begin{gathered}
\frac{\mu_+}{J_+}=\frac{(1+7\lambda^2)\cosh\zeta_--(1+\lambda^2)\cosh(\zeta-\pi\lambda)}{2\mathcal K},\\
\frac{\mu_-}{J_+}=\frac{(1+\lambda^2)\cosh(\zeta+2\pi\lambda)-(1+7\lambda^2)\cosh\zeta}{2\mathcal K},\\
2-g_\pm=\frac{2(1+\lambda^2)\cosh\pi\lambda\,\sinh^2\zeta_\pm}{\mathcal K},\\
\frac{r_0^3\Omega_-}{J_+}=\frac{3\sinh^3\zeta\,\sinh2\pi\lambda}{2\lambda\mathcal K\sinh\zeta_-}.
\end{gathered}
\label{eq:hypforms}
\end{equation}
The hyperbolic forms reduce to the critical
ones as $\lambda\to0$ at fixed $q$.

\textit{Reduction to the unified forms.} Substituting the branch values of
$w$, $q$, $\mathcal B$ and $s_\pi$ into Eqs.~\eqref{eq:deficit}
and~\eqref{eq:Omega-closed} reproduces Eqs.~\eqref{eq:critforms}--\eqref{eq:hypforms}
exactly. On the trigonometric branch the step is the identity
$2(c_+c_-+3\eta^2\mathcal B)=\mathcal F$, which follows from
$c_+c_-=\tfrac12(\cos\pi\eta+\cos2\delta)$ and
$(1-4\eta^2)\mathcal B=\cos\pi\eta$. On the hyperbolic branch it is the equality
of the two expressions for $\mathcal K$, since
$2\sinh\zeta\sinh\zeta_-=\cosh2\beta-\cosh\pi\lambda$. The factor $w^2$
relating $2-g_-$ to $2-g_+$ is $G_-^{-1}$ on every branch.

\section{Uniqueness, angular momentum, limits and numerical checks}
\label{app:proofs}\label{app:komar}\label{app:limits}\label{app:checks}

\subsection{Uniqueness and positivity}

Near either end the homogeneous master equation has solutions behaving as
$|r|^2$ and $1/|r|$. The general solution of Eqs.~\eqref{eq:firstorder}
therefore contains four constants: $C$, two in $p$ and an additive constant in
$\omega$. Decay of $p$ at both ends and $\omega(+\infty)=0$ are three
conditions, which leave the amplitude $C$.

Let $z$ be the difference of two solutions with the same $C$ that satisfy
Eq.~\eqref{eq:bc}. Then $(Gz')'-Uz=0$, with $z=O(|r|^{-1})$ and
$z'=O(r^{-2})$ at both ends. Multiplying by $z$ and integrating,
\begin{equation}
0=\bigl[Gzz'\bigr]_{-\infty}^{\infty}-\int_{-\infty}^{\infty}\bigl(Gz'^2+Uz^2\bigr)\,dr .
\end{equation}
The boundary term vanishes, and $G>0$, $U>0$ force $z\equiv0$. The first
integral and $\omega(+\infty)=0$ then fix $\omega$. The argument uses
neither $Q\ne0$ nor $C\ne0$. For existence, let $p_\pm$ be homogeneous
solutions that decay at the positive and negative ends. If they were
proportional, $p_+$ would be a nonzero homogeneous solution decaying at both
ends, which the identity above excludes. Hence
$\mathcal W=G\,(p_-p_+'-p_-'p_+)$ is a nonzero constant, and
\begin{equation}
\begin{aligned}
p(r)=-\frac{CQ}{\mathcal W}\Bigl[&p_+(r)\int_{-\infty}^{r}p_-\frac{G^2}{S^2}\,dr'\\
&+p_-(r)\int_{r}^{\infty}p_+\frac{G^2}{S^2}\,dr'\Bigr]
\end{aligned}
\end{equation}
is the decaying solution of Eq.~\eqref{eq:master}.

For $Q\ne0$ and $C\ne0$, the function $f=p/(CQ)$ satisfies
\begin{equation}
(Gf')'-Uf=-\frac{G^2}{S^2}<0,\qquad f(\pm\infty)=0 .
\end{equation}
If $f$ were nonpositive somewhere, then, since it vanishes at both ends, it
would have an interior minimum with $f\le0$. There $f'=0$ and $f''\ge0$, so
the left-hand side is nonnegative, which contradicts the sign of the source.
Hence $f>0$ at every finite $r$. Write $f=a_s/|r|+O(r^{-2})$ at the end $s$. If $a_s$
vanished, the leading terms of the equation at large $|r|$, where
$G\to G_s$, $U\simeq2G_s/r^2$ and $G^2/S^2\simeq G_s^2/r^4$, would give
$f=-G_s/(4r^2)+O(|r|^{-3})<0$. Therefore $a_s>0$. With
Eq.~\eqref{eq:asym}, $b_s=CQa_s$ and $Q_sJ_s=-CQ/6$, so
$\gamma_s=6a_s/\sqrt{G_s}>0$, which is Eq.~\eqref{eq:gamma}.

\subsection{Komar angular momentum}

On a slice of constant $t$ the slowly rotating metric has lapse $N=\sqrt G$,
spatial metric $\gamma_{ij}dx^idx^j=G^{-1}(dr^2+S\,d\Omega^2)$ and shift
$\beta^\varphi=-\omega$, with the rotation amplitude absorbed into $\omega$ and
$p$ as in Sec.~\ref{sec:firstorder}. For a stationary metric the extrinsic
curvature is $K_{ij}=(2N)^{-1}(D_i\beta_j+D_j\beta_i)$, and its only independent first-order
component is
\begin{equation}
K_{r\varphi}=\frac{\gamma_{\varphi\varphi}\,\partial_r\beta^\varphi}{2N}
=-\frac{H\,\omega'\sin^2\theta}{2\sqrt G},\qquad H=\frac{S}{G}.
\label{eq:Krphi}
\end{equation}
The angular momentum through the sphere of constant $r$, with unit normal
$n^r=\sqrt G$ pointing toward increasing $r$ and area element
$dA=(S/G)\sin\theta\,d\theta\,d\varphi$, is
\begin{equation}
J_{\rm K}(r)=\frac{1}{8\pi}\oint K_{ij}\,\varphi^in^j\,dA
=-\frac{H^2\omega'}{6},
\label{eq:komarderiv}
\end{equation}
which is Eq.~\eqref{eq:komar}. Since the axial Killing vector
$\varphi^i$ is tangent to the slice, this surface integral is the $3+1$ form of
the Komar integral of $\partial_\varphi$~\cite{Gourgoulhon:2012}. Since
$C=H^2\omega'+4Qp$, the combination $J_{\rm K}-\tfrac23Qp=-C/6$ is the same on
every sphere, and it equals $J_+$ because $p\to0$ as $r\to+\infty$.

\subsection{Kerr--Newman substitution}

The first-order equations are also satisfied by the slowly rotating
Kerr--Newman solution. With $r=R-M$ and $r_0^2=Q^2-M^2$,
\begin{equation}
\begin{gathered}
G=1-\frac{2M}{R}+\frac{Q^2}{R^2},\qquad S=R^2G,\qquad H=R^2,\\
p=-\frac{a_{\rm KN}Q}{R},\qquad
\omega=\frac{a_{\rm KN}(2MR-Q^2)}{R^4},
\end{gathered}
\label{eq:KN}
\end{equation}
with $a_{\rm KN}$ the spin parameter. Substitution into the first-order
equations gives $C=H^2\omega'+4Qp=-6Ma_{\rm KN}$. Hence $J=Ma_{\rm KN}$,
$\mu=a_{\rm KN}Q$ and $g=2$. Equation~\eqref{eq:komarderiv} gives
\begin{equation}
J_{\rm K}(R)=Ma_{\rm KN}-\frac{2a_{\rm KN}Q^2}{3R},\qquad
J_{\rm K}-\frac23Qp=Ma_{\rm KN},
\label{eq:KNkomar}
\end{equation}
in agreement with Eq.~\eqref{eq:komar}. In the notation of
Sec.~\ref{sec:background} this background has no scalar field, $D=0$, so
$a=-1$ and $m^2=q^2-1$; it is the formal member of the family on which the
deficit relation~\eqref{eq:deficit} vanishes, consistent with $g=2$. Its
function $W=R/\sqrt{R^2-2MR+Q^2}$ vanishes at $R=0$, an interior point, so the
member lies outside the regular family.

\subsection{Near-extremal expansions}

On a positive-mouth trajectory at fixed $a=k/r_0^2>-1$ (E1--E3), the expansion
of Eq.~\eqref{eq:firstcorr} continues as
\begin{equation}
\frac{2-g_+}{\epsilon^3}=\frac{16}{27}\biggl[1+K(a)\,\epsilon
-\frac{2\sqrt2\,s_\pi\,\epsilon^{3/2}}{9\sqrt{1+a}\,\mathcal B}
+O(\epsilon^2)\biggr],
\label{eq:fixedDexp}
\end{equation}
where $K(a)$ is given in Eq.~\eqref{eq:firstcorr} and $s_\pi=S_a(\pi)$. For the
critical path ($a=0$, $\mathcal B=1$,
$s_\pi=\pi$) the bracket reads
$1+\tfrac{21}{10}\epsilon-\tfrac{2\sqrt2\pi}{9}\epsilon^{3/2}+O(\epsilon^2)$.
On the symmetric path E4 the
throat identity at $u_{\rm th}=\pi/2$ gives $Q/R_{\rm th}=\eta$, so that
$\epsilon=1-\eta$, and
\begin{equation}
\begin{gathered}
2-g_+=\frac{\pi^2}{3}\epsilon^3+\frac{\pi^2}{2}\epsilon^4
-\frac{\pi^2(\pi^2-6)}{9}\epsilon^5+O(\epsilon^6),\\
\frac{\Omega_-M_+^3}{J_+}=2-\frac{3\pi^2}{2}\epsilon^2+\frac{\pi^2}{3}\epsilon^3+O(\epsilon^4).
\end{gathered}
\label{eq:E4exp}
\end{equation}

\subsection{Rescaled positive exterior}

On E1--E4, $m=M_+/r_0\to\infty$. At fixed $\chi=r/M_+>0$ one has
$u=\arctan(r_0/r)=1/(m\chi)+O(m^{-3})$, so Eq.~\eqref{eq:Wunified} gives
\begin{equation}
\begin{gathered}
W=C_a(u)+mS_a(u)=1+\frac1\chi+O(m^{-2}),\\
\frac{S}{M_+^2}=\chi^2+\frac{1}{m^2}.
\end{gathered}
\label{eq:RNsteps}
\end{equation}
With $\tau=t/M_+$ and $G=W^{-2}$, the rescaled static background metric
\begin{equation}
\frac{ds^2}{M_+^2}=-\frac{d\tau^2}{W^2}+W^2\Bigl(d\chi^2+\frac{S}{M_+^2}\,d\Omega^2\Bigr)
\end{equation}
tends to Eq.~\eqref{eq:RNlimit}.

\subsection{Numerical checks}

The numerical results in the figures and tables use $r_0=1$. The closed-form checks use a working
precision of at least 40 digits. For the configurations of
Fig.~\ref{fig:profiles}, with $C=1$, substitution into both first-order
equations and the first integral gives residuals below $10^{-35}$. The
deficit relation~\eqref{eq:deficit} agrees with the branch formulas of
Appendix~\ref{app:branches} to better than $10^{-25}$ at the 4018 sampled
configurations of Fig.~\ref{fig:gmaps}. The closed form of
$r_0^3\Omega_-/J_+$ also agrees with direct quadrature of
Eq.~\eqref{eq:Omega-integral}, with a normalized discrepancy below $10^{-40}$ at
the tested points.
Independent Python calculations with arbitrary-precision arithmetic confirm
selected numerical values underlying every figure and both tables. Across
the 18 hyperbolic boundary-value runs, the largest recorded discrepancy in
$p$ at the sampled radial points is $1.7\times10^{-9}$.

\begin{acknowledgments}
This work was supported by the Deanship of Scientific Research, Vice Presidency for Graduate Studies and Scientific Research, King Faisal University, Saudi Arabia (Funding Grant No: KFU266635). 

\end{acknowledgments}

\section*{Data availability}
The numerical data underlying the figures and tables in this study are
available from the corresponding author upon reasonable request.

\bibliography{references}

\end{document}